\documentclass[namedreferences]{solarphysics}

\usepackage[hyperref,optionalrh,solaromanenum]{spr-sola-addons} 
\usepackage{graphicx}        
\usepackage{color}           
\usepackage{breakurl}        
\usepackage{lmodern} 
\usepackage{longtable}
\usepackage[utf8]{inputenc} 
\usepackage[T1]{fontenc} 

\DeclareGraphicsExtensions{.eps}

\newcommand{\caii}{Ca\,{\sc ii}\,K}
\newcommand{\ha}{H$\alpha$}

\begin{document}

\begin{article}

\begin{opening}

\title{Kanzelh\"ohe Observatory: instruments, data processing and data products}

\author[addressref={aff1},email={werner.poetzi@uni-graz.at}]{\inits{W.}
        \fnm{Werner}~\lnm{P\"otzi}\orcid{0000-0003-4811-0543}}
\author[addressref={aff1,aff2},email={astrid.veronig@uni-graz.at}]
       {\inits{A.M.}\fnm{Astrid}~\lnm{Veronig}\orcid{0000-0003-2073-002X}}
\author[addressref={aff2},email={robert.jarolim@uni-graz.at}]{\inits{R.}
       \fnm{Robert}~\lnm{Jarolim}\orcid{0000-0002-9309-2981}}
\author[addressref={aff3},email={J.RodriquezGomez@skoltech.ru}]{\inits{J.M.}
       \fnm{Jenny~Marcela}~\lnm{Rodr{\'\i}guez~G\'{o}mez}\orcid{0000-0002-9860-5096}}
\author[addressref={aff3},email={T.Podladchikova@skoltech.ru}]{\inits{T.}
       \fnm{Tatiana}~\lnm{Podladchikova}\orcid{0000-0002-9189-1579}}
\author[addressref={aff1},email={dietmar.baumgartner@uni-graz.at}]{\inits{D.J.}
        \fnm{Dietmar}~\lnm{Baumgartner}\orcid{0000-0002-9794-0458}}
\author[addressref={aff1},email={heinrich.freislich@uni-graz.at}]{\inits{H.}
        \fnm{Heinrich}~\lnm{Freislich}\orcid{}}
\author[addressref={aff1},email={heinz.strutzmann@uni-graz.at}]{\inits{H.}
        \fnm{Heinz}~\lnm{Strutzmann}\orcid{}}

\address[id=aff1]{Kanzelh\"ohe Observatory for Solar and Environmental 
   Research, University of Graz, Treffen am Ossiachersee, Austria}
\address[id=aff2]{Institute of Physics/IGAM, University of Graz, Graz, Austria}
\address[id=aff3]{Skolkovo Institute of Science and Technology, Moscow, 
   Russia}

\runningauthor{P\"otzi {\it et al.}}
\runningtitle{KSO data processing and data products}

\begin{abstract}
Kanzelh\"ohe Observatory for Solar and Environmental Research (KSO) 
of the University of Graz (Austria) is in continuous operation since its 
foundation in 1943. Since the beginning its main task was the regular 
observation of the Sun in full disc. In this 
long time span covering almost seven solar cycles, a substantial amount of data
was collected, which is made available online. In this paper we  
describe the separate processing steps from data acquisition to
high level products for the different observing wavelengths. First of all
we work out in detail the quality classification, which is important for
further processing of the raw images. We show how we construct 
centre-to-limb variation (CLV) profiles and how we remove large scale 
intensity variations produced by the telescope optics in order
to get images with uniform intensity and contrast. 
Another important point is an overview
of the different data products from raw images to high contrast
images with heliographic grids overlaid. As the data products are
accessible via different sources we also present how to get information 
about the availability and how to obtain these data.
Finally, in an appendix, we describe in detail the information in the 
FITS headers, the file naming and the data hierarchy.
\end{abstract}
\end{opening}


\section{Introduction}

Kanzelh\"ohe Observatory for Solar and Environmental Research (KSO) 
of the University of Graz is located on a mountain ridge at an altitude 
of 1526 m a.s.l. near Villach in southern Austria. The observatory 
was founded in 1943 as part of a network of solar observatories operated 
by the German Airforce \citep{Poetzi2016,Jungmeier2017}. The discovery of 
M\"ogel and Dellinger \citep{Dellinger1935} that solar flares can cause 
ionospheric disturbances by a sudden increase of the ionisation of the 
D-layer in the ionosphere, such that blackouts of radio communications 
can occur, made it necessary to investigate such phenomena to arrive at 
a better understanding and predictions of these effects \citep{Seiler2007}.

\begin{figure} 
 \centerline{\includegraphics[width=1\textwidth,clip=]{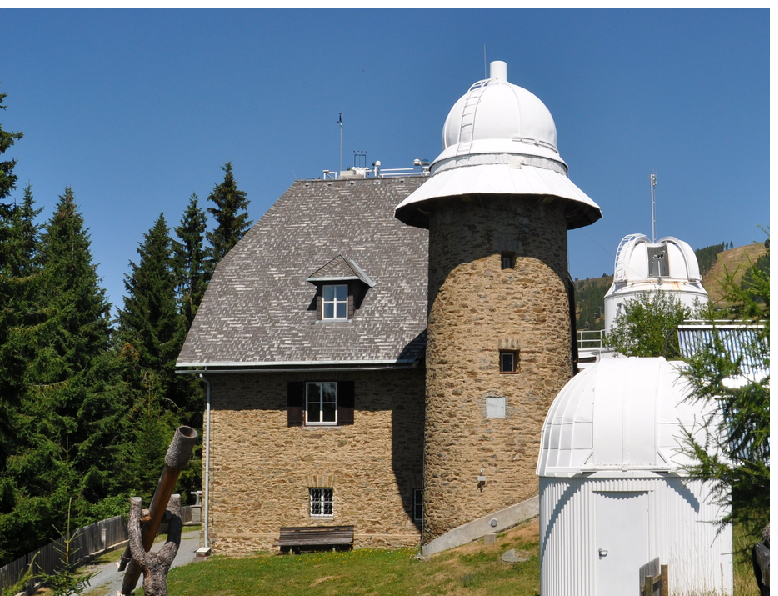}}
 \caption{Kanzelh\"ohe Observatory today: The heliostat was installed 
 in the southern tower (leftmost) in the northern tower (open dome in background) }
 a small telescope for testing purposes replaced the coronagraph after 1948. Since 1973
 all observations are performed in the northern tower.
 \label{fig:obs}
\end{figure}

Solar observations at KSO date back to 1943. In the early years of the 
observatory, most observations were performed using a heliostat guiding 
the light beam down to a laboratory in the basement, where a projection device 
for sunspot drawings and a spectroheliograph \citep{Siedentopf1940,Comper1958} 
were installed. In a second tower a coronagraph was operated \citep{Comper1957},
which was transferred in 1948 to the top of the Gerlitzen mountain (1\,900 m a.s.l).
First observations in the \ha\ spectral line were made visually at the spectroheliograph. 
In 1958, a Zeiss-Lyot monochromator acquired on occasion of the International 
Geophysical Year made photographic observations in the \ha\ line possible. 
In addition, a piggy-back mounted telescope on the coronagraph was used to 
record the photosphere on glass plates. A more detailed description about 
the early observations at KSO can be found in \cite{Kuiper1946}.
Corona observations were suspended in 1964 as the seeing conditions 
on top of the mountain worsened. In 1973 all 
observations were transferred to the northern tower (Figure \ref{fig:obs}) 
offering better seeing conditions \citep{Poetzi2016}.

In pre-digital times, {\it i.e.} up to the year 2000, observing was a 
combination of ``manual'' and ``visual'' work. Photographs were made on 
glass plates (later on sheet film and 36\,mm film), which had to be manually exposed 
and afterwards developed. Observations at the coronagraph
were on the one hand noted on special form sheets describing the 
intensity in the green (5303\,\AA) and red (6874\,\AA) corona lines 
({\it cf.} \opencite{Waldmeier1942}) and on the other hand also photographs
of the corona intensity in \ha\ (FWHM=20\,\AA) were taken \citep{Comper1957}. 
White-light observations on photographic material date back to 1943, 
images were recorded irregularly,
mostly on occasion of big sunspots. From 1989 on every day at least 
three images were taken using the photoheliograph in the northern tower
\citep{Pettauer1990}. Patrol observations of \ha\ images started 
in 1958 on an irregular basis and were already automatized in 1973 at a
cadence of one image every 4 minutes on 35\,mm film rolls. Large parts of the 
data have been digitized like the white-light images on sheet film from 1989 to 2007 
\citep{Poetzi2010} and the \ha\ film rolls from 1973 to 2000 
\citep{Poetzi2007}. Digitising the corona images and the \ha\ images
before 1973 is not planned as the effort is too high and the quality of the
data is often very low. Digitizing the white-light images on glass
plates between 1943 and 1970 is planned in the future.
Sunspot drawings have always been made by hand and they will not be subject 
to automatized procedures in order to contain long-term consistency, as 
is explained in detail by \cite{Clette2014}. Nonetheless all drawings are 
scanned immediately and published in our archive. In 2010 full disc observations 
in the \caii\ line widened out the monitoring of the solar activity.

KSO data are used for a variety of scientific research topics, due to the 
data reaching back over several cycles, their continuous and high data 
quality as well as for being freely available to the community via the 
KSO data archive (\url{https://cesar.kso.ac.at}). The main research 
topics for which KSO data are studied are the following. \ha\ data are in 
particular used for the study of energetic events such as solar flares 
\citep[{\it e.g.},][]{Wang2003,Veronig2006a,Vrsnak2006a, Miklenic2007, Poetzi2015,
Thalmann2015, Joshi2017, Temmer2017, Hinterreiter2018, Tschernitz2018}, 
filament eruptions and dynamics \citep[{\it e.g.\rm}][]{Maricic2004,Moestl2009, 
Zuccarello2012, Su2012, Filippov2013, Yardley2016, Aggarwal2018}, 
Moreton waves \citep[{\it e.g.},][]{Pohjolainen2001,Warmuth2001, Vrsnak2006b, Veronig2006b} 
as well as for long-term evolution of the distributions of prominences 
and polar crown filaments  \citep[{\it e.g.},][]{Rybak2011,Xu2018, Xu2021, Chatterjee2020}. 
Sunspot drawings are used for the derivation of (hemispheric) sunspot numbers 
and solar cycle studies \citep[{\it e.g.},][]{Temmer2002, Temmer2006, Clette2014, 
Clette2016, Poetzi2016, Chatzistergos2017, Chowdhury2019,Veronig2021}. 
White light images and sunspot drawings are used  for deriving solar 
differential rotation rates and meridional flows 
\citep[{\it e.g.}\,][]{Hanslmeier_1986,Lustig_1987,Poljancic2017,Ruzdjak2018} 
as well as for sunspot 
catalogs \citep{Baranyi2016}. \caii\ images are used for long-term solar 
irradiance investigations \citep[{\it e.g.},][]{Chatzistergos2020} and flare studies 
\citep[{\it e.g.},][]{Veronig2015, Sindhuja2019}.


\section{Instruments and Data acquisition} 

\subsection{Instruments}

\begin{figure}    
   \centerline{\includegraphics[width=1.0\textwidth,clip=0]{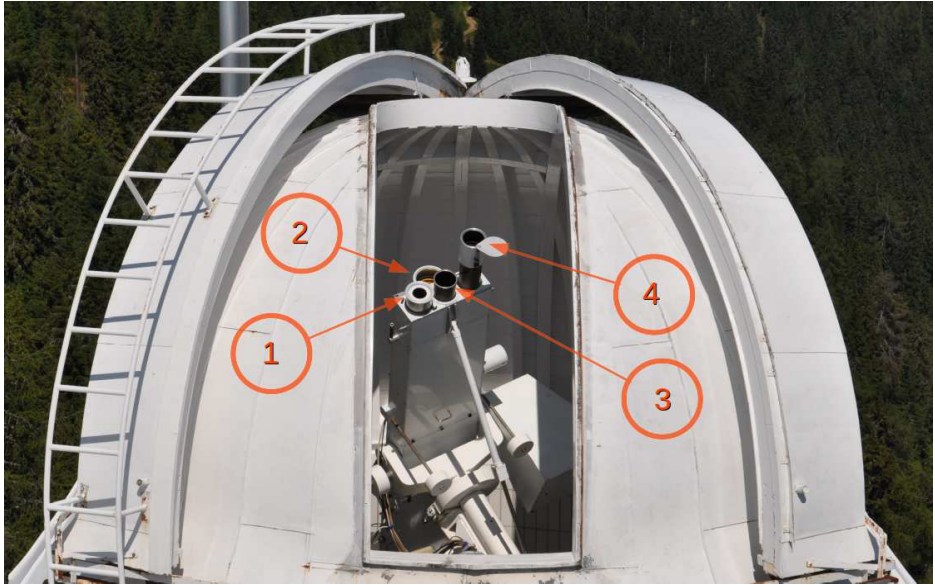}}
   \caption{The KSO solar patrol instrument in the northern tower carries four telescopes:
   (1) - Ca\,{\sc ii}\,K, (2) - White-light, (3) - \ha, (4) - Drawing device. The slit
   of the dome moves automatically with the telescope.}
   \label{fig:UEWI}
 \end{figure}

The main instrument at KSO is the patrol instrument, which is in operation 
since 1973 in the northern tower. This instrument 
carries four telescopes (Figure \ref{fig:UEWI}) on a parallactic mounting:
\begin{itemize}
 \item [(1)] {\bf \caii:} \\A refractor with a diameter of 100\,mm and a focal 
     length of 1\,500\,mm. The system is protected by a dielectric energy
     reflection filter. The main filter was originally a DayStar etalon,
     which degraded in 2012 and was replaced by a Lunt B1800 Ca-K module 
     in 2013. As the specifications required a F/20 configuration
     in order to guarantee the designed bandwidth \citep{Polanec2011} the
     aperture front lens was reduced to 70\,mm.
 \item [(2)] {\bf White-light:}\\ A refractor with an objective lens of
     130\,mm and a focal length of 2000\,mm. The broadband filter with a
     FWHM of 100\,\AA\ is centred in the green range of the visual electromagnetic
     spectrum
     at 5\,460\,\AA\ \cite[]{Otruba2008}. The telescope is based on the 
     Kanzelhöhe Photoheliograph (PhoKa) which went into operation in 1989 
     and was described in detail in \citet{Pettauer1990}.
     In order to compensate for thermal expansion the camera position 
     is adapted each day.
 \item [(3)] {\bf \ha:}\\ A refractor with 100\,mm diameter and a focal length of 
    1\,950\,mm. The objective lens is coated with gold serving as heat filter and
    the Zeiss Lyot \ha\ filter with a central wavelength of 6\,562.8\,\AA\
    and a FWHM of 0.7\,\AA\ is placed behind a broadband \ha\ pre-filter
     \citep{Otruba2003,Poetzi2015}. A beam-splitter provides the possibility 
     to attach a second camera, which can be used for, {\it e.g.}, 
     ultra-high cadence observations with several images per second 
     additional to the standard patrol observations.
 \item [(4)] {\bf Drawing Device:}\\ The drawing device is mounted near the declination
     axis of the telescope, which makes it comfortable to produce the
     drawing on the projected image (which has a diameter of 25\,cm) 
     and also minimizes the forces applied to the telescope by the drawing
     procedure itself. The drawing is used for obtaining the relative 
     sunspot number, which is sent to SILSO (Sunspot Index and Long-term 
     Solar Observations\footnote{\url{http://sidc.oma.be/silso/}}) 
     World Data Center at the Royal Observatory of Belgium for calculating the 
     International Sunspot Number (ISN),
     more details can be found in \citet{Poetzi2016}.
\end{itemize}

An overview of the telescope configurations is given in Table \ref{tbl:patrol}.
In the current configuration  the same CCD camera type (2048$\times$2048 pixels
with 4\,096 intensity levels) is used for data acquisition  (further 
technical details  on the camera specification can be found in Appendix \ref{camera}). 
Identical grabbing software and almost the same processing software is implemented for all 
observation types. This configuration reduces the workload in 
handling, maintaining and keeping up to date. As a measure of precaution 
we have acquired additional CCD cameras of the same type for immediate 
replacement if necessary.

\begin{table}
\caption{The  current and former configurations of the telescopes on the 
KSO solar patrol instrument:}
\label{tbl:patrol}
\begin{tabular}{lccc}\hline
 \multicolumn{4}{c}current configurations (2021) \\ \hline
                    & \ha     & White-light    &  Ca\,{\sc ii}\,K \\ \hline
 aperture           & 100\,mm       & 130\,mm        &  70mm \\
 focal length       & 2000\,mm      & 1\,950\,mm     &  1\,500\,mm \\
 central wavelength & 6\,562.8 \AA  & 5\,450 \AA     &  3\,933.7\AA \\
 FWHM               & 0.7 \AA       & 100 \AA        &  20 \AA \\
 CCD camera         & \multicolumn{3}{c}{all telescopes: JAI Pulnix RM-4200GE} \\
 image size         & \multicolumn{3}{c}{2\,048 $\times$ 2\,048 pixel} \\
 bit depth          & \multicolumn{3}{c}{12bit} \\
 observing cadence  & 10 images/min & 3 images/min & 10 images/min \\
 exposure time      & 1.5 - 25 ms   & 2.2 - 25 ms & 1.5 - 35 ms \\
 operating since    & June 2008     & August 2015    & July 2010 \\ \hline
 \multicolumn{4}{c}{former configurations} \\ \hline
 CCD camera         & Pulnix TM-1010 & JAI Pulnix TM-4100CL & \\
 image size         & 1\,000 $\times$ 1\,1012 pixel & 2\,048 $\times$ 2\,048 pixel & \\
 bit depth          & 10bit  & 10bit  & \\
 observing cadence  & 10 images/min & 1 image/min &  \\
 exposure time      & 2.5 - 33 ms   & 3.5 - 25 ms &  \\
 operating since    & July 2005  & July 2007    &  \\ \\
 CCD camera         & Pulnix TM-1001 &  & \\
 image size         & 1\,008 $\times$ 1\,016 pixel & & \\
 bit depth          & 8bit & & \\
 observing cadence  & 1 image every 100 sec &  &  \\
 exposure time      & 3.5 - 66 ms   & &  \\
 operating since    & December 1998  &   &  \\
\hline 
\end{tabular}
\end{table}

\subsection{Data acquisition}

\begin{figure} 
 \centerline{\includegraphics[width=1\textwidth,clip=]{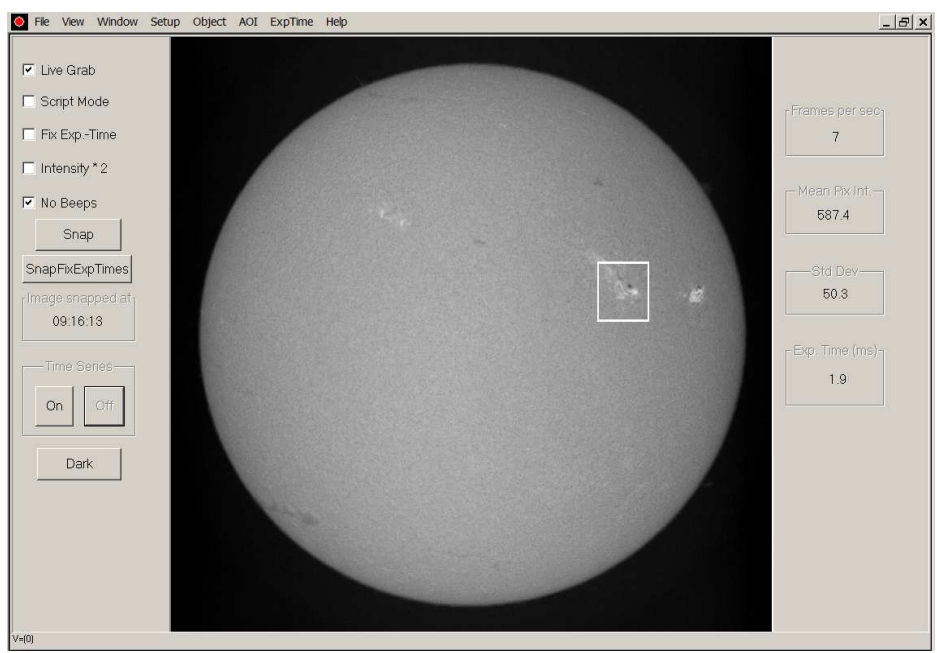}}
 \caption{The data acquisition interface for the \ha\ camera
 (26~May~2021), it is the same for the other cameras. In standard patrol
 observation mode the user has just to click the ``On'' button and the
 observation program runs automatically. The white box indicates the 
 selected AOI.}
 \label{fig:interf}
\end{figure}

The data acquisition software is written in C++ and makes use of commercial
camera control libraries (Common Vision Blox from Stemmer Imaging). The
user interface (Figure \ref{fig:interf}) is kept as simple as possible but also offers many 
additional features, which are not used in standard patrol mode.
The observer marks a rectangular box on the solar disc: 
during enhanced solar activity this box is placed over an
active region and in case of quiet solar conditions the box is
placed at solar disc centre. 
This AOI (area of interest) is used for calculating the exposure time 
and the image contrast. The mean brightness in this AOI is set to a fixed 
value (600 counts for \ha\ in order to avoid saturation of the CCD 
chip in case of bright flares, 2\,000 for white-light and 1\,000 for \caii). 
Based on these pre-defined mean values, the exposure time for the camera 
is calculated. As the camera is much faster (6 to 7 images per second) 
than the typical image cadence (6 seconds) the system can use frame 
selection and profits from very short moments of better seeing 
conditions \citep{Scharmer1989}. For this purpose the image contrast 
in the AOI is calculated, {\it i.e.}, the camera selects from a sequence 
of about 10 images the one that has the highest {\it rms} contrast.

Before and after the daily observations, when the objective lens is protected with
the dust cover, a dark current image is made. For this purpose the exposure
time is fixed to 3\,msec, which corresponds to an average exposure time in
patrol mode. Dark current images during the observation time are not 
possible as there always falls light onto the CCD chip, since the cameras 
are not equipped with a mechanical shutter. The button ``SnapFixExpTimes'' 
starts a series of images at 5\,msec, 20\,msec and 50\,msec exposure time. 
It is used especially for the \ha\ camera in order to get an image of the 
lower corona in \ha\ light by overexposing the solar disc. These images 
are used as backup and additional data for the solar prominence catalogue
of the Lomnick\'y \v{S}t\'it Observatory \citep{Rybak2011}.
All images are stored directly in a live archive on a raid system as raw 
FITS files and JPEG files for a quick inspection.

\section{Data processing}

\subsection{Data pipeline}

\begin{figure} 
 \centerline{\includegraphics[width=1\textwidth,clip=]{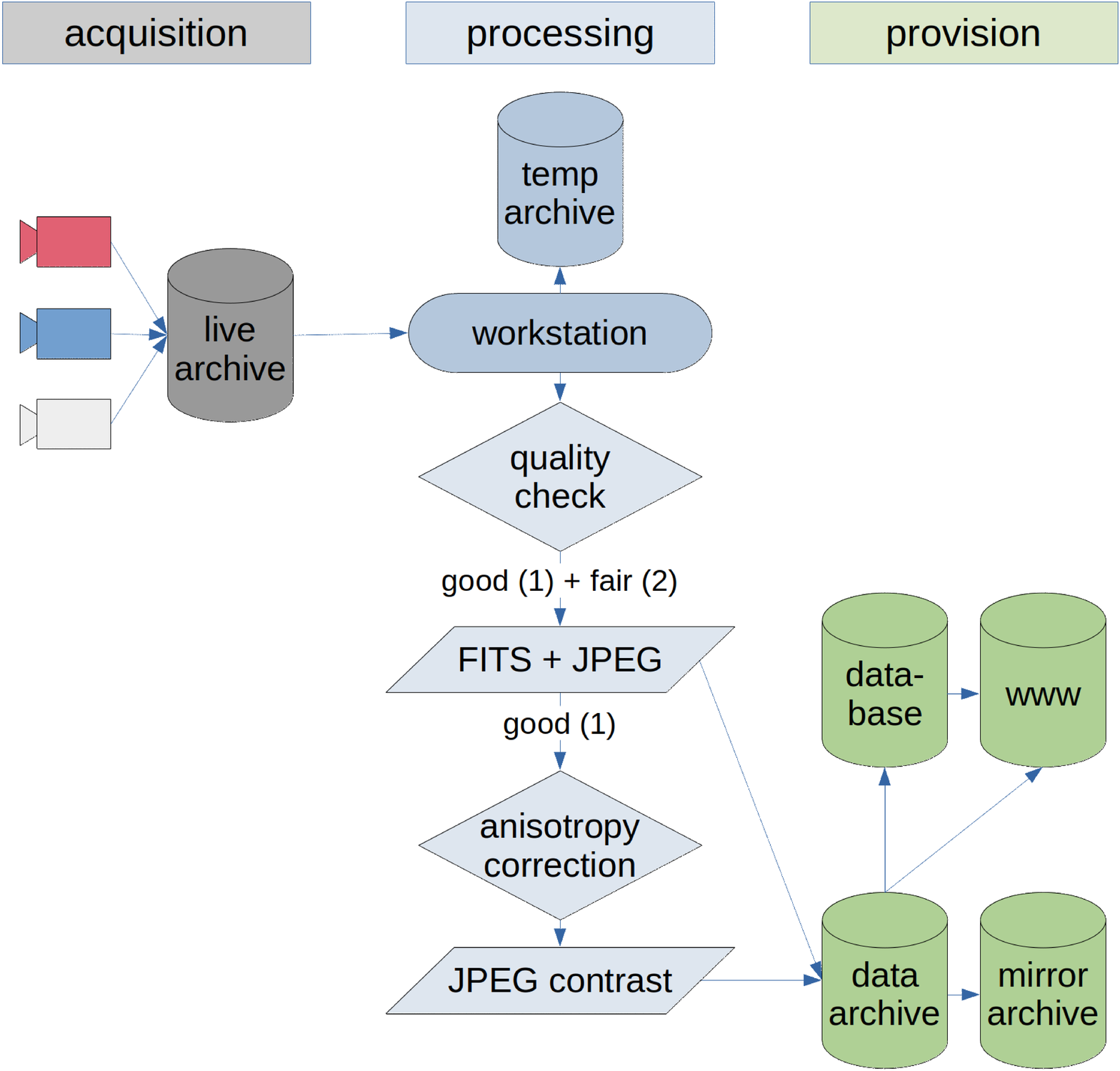}}
 \caption{Flow diagram of the data pipeline for solar images at KSO. Each of the three
 branches -- acquisition, processing and provision -- is running on a
 separate system. All systems work in parallel, the whole process takes 
 not more than 5 seconds for an individual image from acquisition to provision.
 }
 \label{fig:pipel}
\end{figure}

Figure \ref{fig:pipel} shows the data pipeline for images from 
their recording at the CCD camera
to the main archive. The whole process can be divided into three steps: 
data acquisition, data processing and data provision.
Each step is performed by a separate workstation, which work in parallel. 
The whole process from data acquisition to the provision of the processed 
data takes about 5 seconds per image. \ha\ images undergo a second 
step of processing (FITS files of quality 1) in order to automatically detect and
characterize flares and filaments in the real-time observations. 
This procedure is elaborated in detail in \cite{Poetzi2015} and evaluated 
in \cite{Poetzi2018}. 

In the following, we describe the standard 
processing pipeline, which every image grabbed by one of the CCD cameras 
passes through. In order to keep the system simple and to reduce maintenance and 
work load for updates all following steps are performed by almost the 
same script that is executed three times in parallel, one instance for 
each of the 3 CCD cameras. This script calls individual programs, which 
process the images or manage the database. The scripts are started before 
sunrise and just wait for new images appearing in the live archive for 
processing until they stop after sunset.

{\bf Step 1:}
Before processing, the observed image is transferred from the
live archive to the temporary archive, which can hold up to nearly 
one year of the raw observation data.
In order to free space for 
new images, successively the oldest ones in this temporary archive are deleted.
Normally in the live archive
there is only the latest image. If there is some interruption or delay 
in the processing pipeline, a larger amount of images can accumulate 
there. But also in this case only the latest image is processed and 
transferred in order to prevent any delay in data provision. 
All images that remain on the live archive until the end of the 
observation day are processed later.
The temporary archive is used as an emergency backup, 
in case the processing fails due to some unexpected 
reason or if even images of low quality can be of interest for inspection of some
extraordinary solar events. It can also be of interest in developing 
algorithms for the parametrization of image quality or classification 
purposes \citep{Jarolim2020}, see Section \ref{subimqual}. 

{\bf Step 2:}
A crucial step is the quality check (described in detail in Section 
\ref{subimqual}), that selects the images that are processed and 
stored in the main data archive. This 
procedure should be fast and deliver stable results independently of the 
solar cycle phase or time of day. The images are categorized as good 
(quality=1), fair (2) and bad. Images of quality 1 are 
suitable for all further automatic processing steps, whereas quality 2 images 
can still be used for visual inspection.


{\bf Step 3:}
Images of quality 2 (fair image quality, suitable for visual inspections)
and of quality 1 (good quality) get a an updated FITS header containing all 
relevant image information, see Appendix \ref{FITS}) and a normal contrast 
JPEG image is produced for real time display. The JPEG image is centred, 
contrast enhanced by unsharp masking and contains logos, date and time stamp 
and is overlaid with a heliographic grid.
The FITS file and the JPEG file are stored in the data archive.

{\bf Step 4:}
From images of quality 1 additionally high contrast JPEG images are 
generated, {\it i.e.} they are corrected for global an\-iso\-tropies 
in order to provide a homogeneous contrast across the solar disc 
(details in Section \ref{inhomoCorr}). Like normal contrast images, 
they are overlaid with a heliographic grid, they contain
logos and a date and time stamp. These images are also stored 
in the data archive.

Dark current is not subtracted from the images, This effect cannot be seen in the 
JPEG images as the dark current correponds to a very low level of additional 
noise, which is removed by the JPEG compression algorithm. Flat field 
is also not applied for reasons explained in Section \ref{subFF}.

\subsection{Image quality detection}\label{subimqual}

In this section we describe the details of the image quality determination, 
which is in operation at KSO. The quality estimate makes use of a variety 
of properties of the images, which allow us to characterize the quality 
on both global and local scales:
\begin{itemize}
    \item exposure time
    \item mean intensity in AOI
    \item solar limb and radius determination, see \ref{detlimb}
    \item large-scale image inhomogeneities, see \ref{iminho}
    \item image sharpness, see \ref{detsharp}
\end{itemize}
Since the exposure time is automatically controlled and set in a way 
that the AOI has a pre-defined mean count value, it is sensitive 
to clouds and to a low elevation of the Sun above the horizon.
The mean intensity in the AOI can vary when clouds move very fast across the solar 
disc, in such cases the exposure time adjustment can be too slow, and the 
image can be under- or overexposed. 

\begin{figure} 
 \centerline{\includegraphics[width=1\textwidth,clip=]{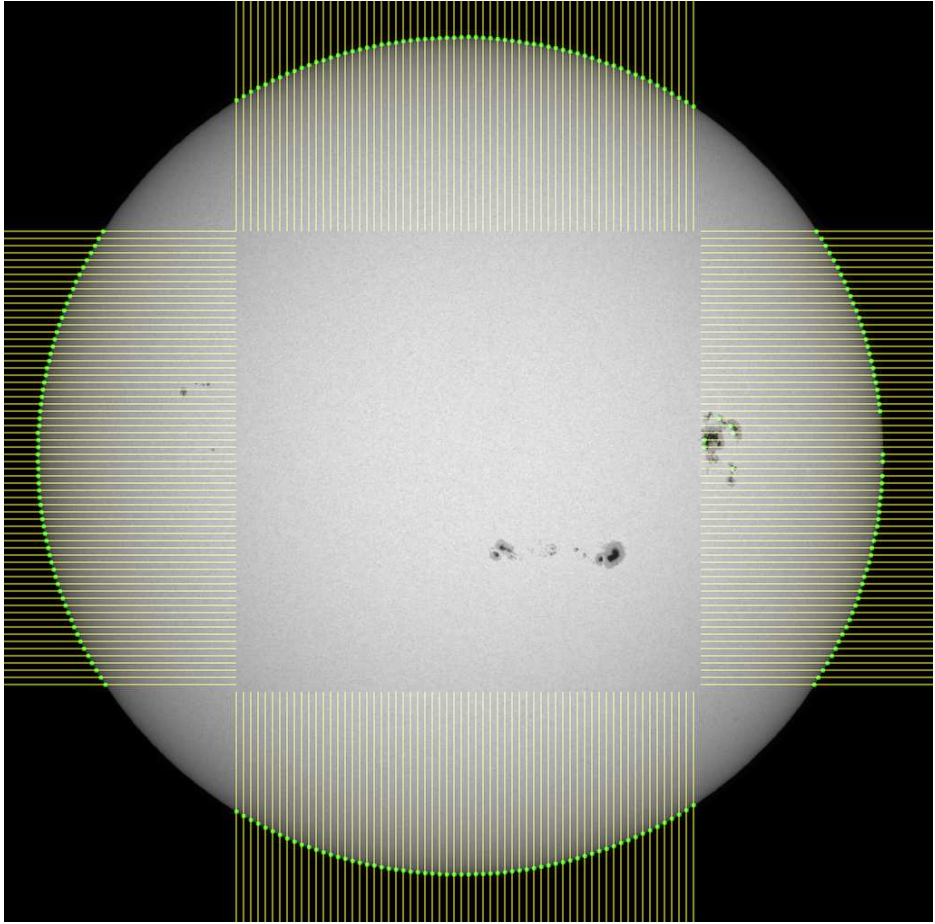}}
 \caption{Illustration of the determination of the solar limb and centre.
 The maximum gradient (green dots) of the intensity in 64  profiles for 
 each direction is calculated. A circle fit through these 
 points gives the radius and the centre coordinates of the solar disc.
 In this white-light image from 6~September~2017 large sunspots lead to
 some outliers in the first step of the limb detection, which are 
 automatically removed in the following iterative steps.}
 \label{fig:limbp}
\end{figure}

\subsubsection{Detection of solar limb and radius}\label{detlimb}

This procedure is an enhanced version of the method published by 
\cite{Veronig2000} and is based on the limb detection by finding the 
limb points as the maximum gradient of intensity profiles in $x$ and $y$ 
direction. To get stable results the profiles are smoothed over 7 pixels 
before the gradient is applied via a morphological operation.
For reducing the calculation time only 64 profiles from each direction 
are generated (see Figure \ref{fig:limbp}). 
Through these limb points a circle is fitted using Taubins 
method \citep{Taubin1991}. This method is very stable and fast and 
it needs only a limited number of points for the fit. 
The circle fit tries to minimize 
\begin{equation}
  \Omega = \sum_{i=1}^{n}d_i^2
\end{equation}
where
\begin{equation}
  d_i = \sqrt{(x_i-a)^2+(y_i-b)^2}-R_{\odot}
\end{equation}
which is the geometric distance from each detected limb point ($x_i$,$y_i$) on
the solar limb to the fitting circle whose centre and radius are ($a$,$b$) 
and $R_{\odot}$, respectively. As a measure for the quality of the circle fit we
obtain the value $\sigma$ which is the rms of $d_i$. In order to improve 
the circle fit and to obtain a robust solution, outliers are iteratively 
removed. {\it I.e.,} in the first iteration limb points with a distance 
from the fitted circle $d\ge R_{\odot}/30$ are excluded and a new circle fit is
performed. If $\sigma$ is still $\ge 2$ this procedure is repeated with 
$d \ge R_{\odot}/100$ and if necessary also with $d \ge R_{\odot}/200$. 

\begin{figure} 
 \centerline{\includegraphics[width=0.5\textwidth,clip=]{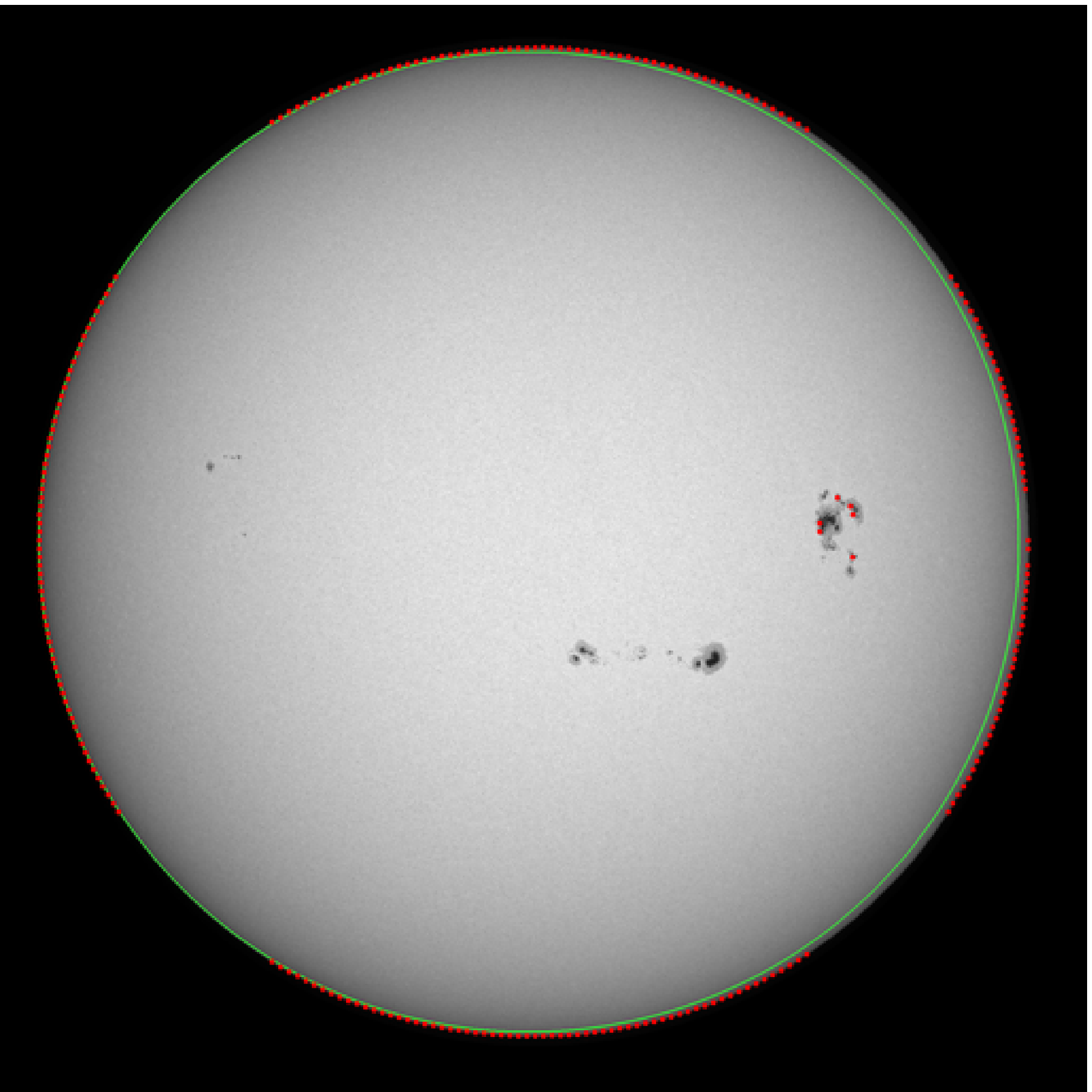}
 \includegraphics[width=0.5\textwidth,clip=]{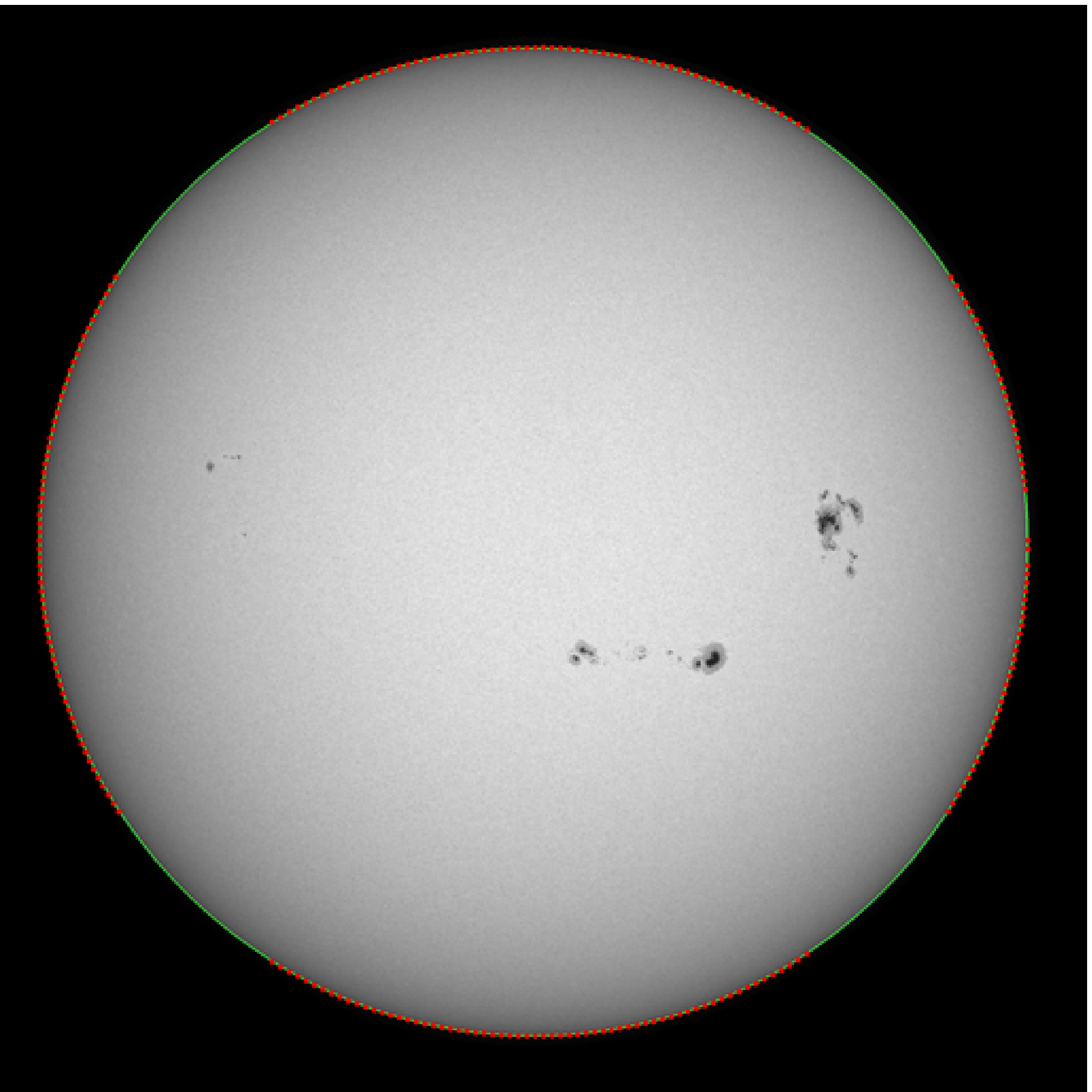}}
 \centerline{\includegraphics[width=0.5\textwidth,clip=]{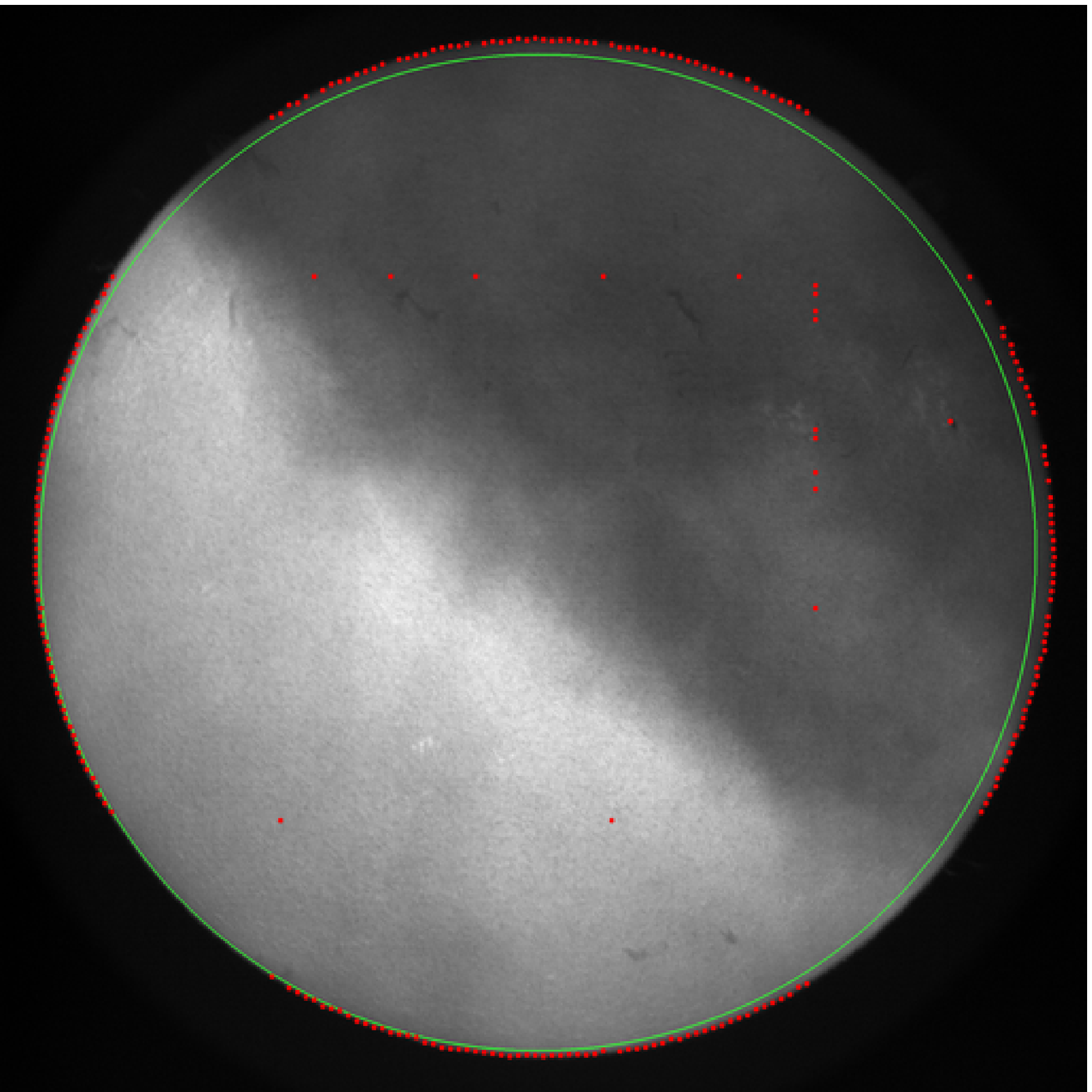}
 \includegraphics[width=0.5\textwidth,clip=]{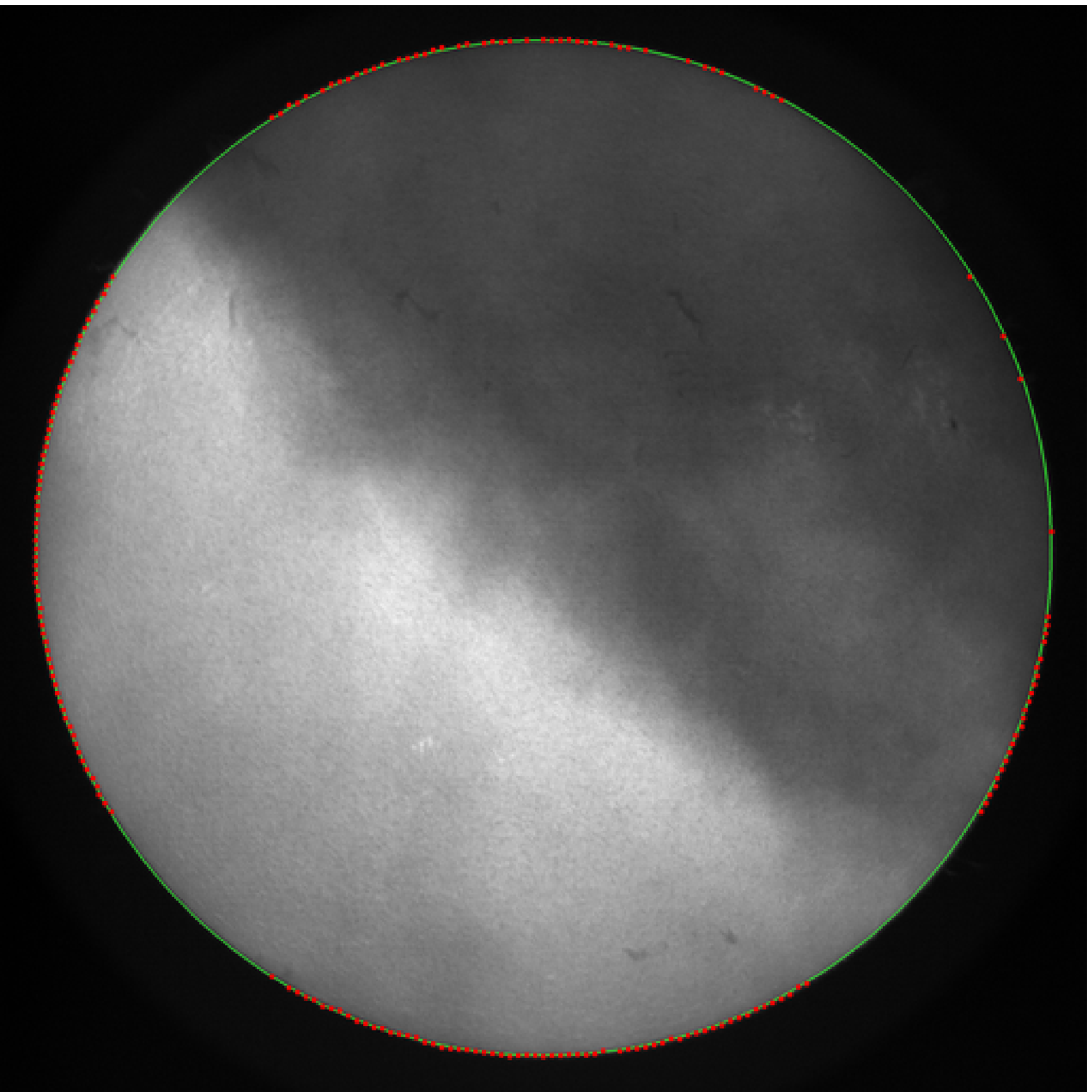}}
 \caption{Illustration of the determination of the solar limb and centre.
 Detected limb points (red dots) and limb fit (green circle) obtained after 
 the first (left panels) and final (right panels) iterations.
 In the first row a white-light image from 6~September~2017 
 at good observation conditions shows already after one iteration 
 a good fit for the solar limb.
 In the second row clouds disturbed the \ha\ observations on 31~December~2020 
 and therefore three iterations where necessary to obtain a good circle fit.}
 \label{fig:limb}
\end{figure}

In Figure \ref{fig:limb} two examples demonstrate the workflow of this
process. The white-light image at good observing conditions in the first 
row needs just one iteration for finding sufficiently precise solar disc parameters, 
the outliers near the large sunspot group (red dots) are removed immediately. 
In the second row an \ha\ image is strongly disturbed by clouds. 
In this case all three iteration 
steps are needed in order to obtain a good circle fit through the limb. 
As the profiles used for detection of the limb points go from the image 
border one fourth into the image ({\it cf.} Figure \ref{fig:limbp}), 
some of the detected points are aligned in almost a straight line
(red dots in left image of the second row in Figure \ref{fig:limb}).
These ``outliers'' are responsible for the bad circle fitting 
(overplotted green circle). Therefore in the next iterations outliers are 
removed and the circle fit is recalculated. The right image shows the 
result obtained in the last step, where the circle fit (indicated as 
green line)  is within the required limits ($\sigma \le 2$). 
We note that such image is not further 
processed. This is an example to demonstrate how robust this method is. 
We note that the result of this limb finding procedure is also used to  
fine-tuning the telescope guider. Thanks to this robustness, the telescope 
position can be maintained accurately to the centre of the solar disc also 
during cloudy periods.

\begin{figure} 
 \centerline{\includegraphics[width=1\textwidth,clip=]{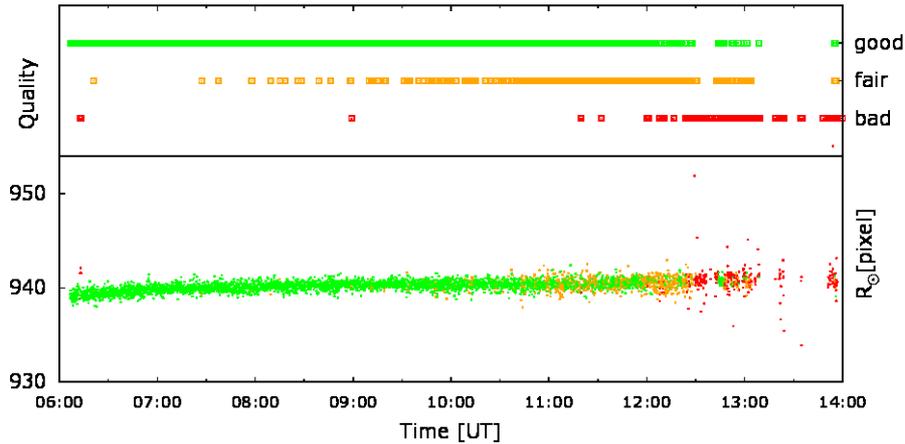}}
 \caption{\ha\ image quality and solar radius determination for a full 
 observation day (11~April~2020). Top panel: The observed \ha\ filtergrams 
 are colour coded by the determined image quality. Bottom panel:
 The radius determinations are coloured according to the corresponding image
 quality. This is a typical seeing situation during a fine sunny day at KSO, 
  where the seeing conditions become worse during the day and cumulus clouds evolve
 in the afternoon. The data series plotted contains a total of 4204 images.}
 \label{fig:qualrad}
\end{figure}

Figure \ref{fig:qualrad} shows the determined radius and the \ha\ image quality 
for a complete observation day for the \ha\ telescope (11~April~2020). 
In total 4\,204 \ha\ images were recorded on this day: 3\,098 of quality 
class 1, 738 of class 2 and 375 of quality 3. One can see from the 
plot, that the observations get worse over the day. The observation 
conditions are commonly the best in the morning. In case of fine and sunny
weather, over the day the rising air causes turbulences that decrease the seeing 
conditions. In the afternoon occurring cumulus clouds interrupt the observations, 
in most cases these clouds dissolve until the evening. The determined radius 
of the solar disc (bottom panel) reveals in general very stable results, 
with an average $R_{\odot} = 940.16 \pm 0.52$ pixels (determined from all 
images of quality class 1). The radius values in the morning and later 
in the  afternoon are somewhat smaller due to refraction effects and less 
image motion. As one can also see, the radius distribution becomes wider when the image 
quality decreases.

\subsubsection{Parametrization of large-scale image inhomogeneities}\label{iminho}

\begin{figure} 
 \centerline{\includegraphics[width=1.0\textwidth,clip=]{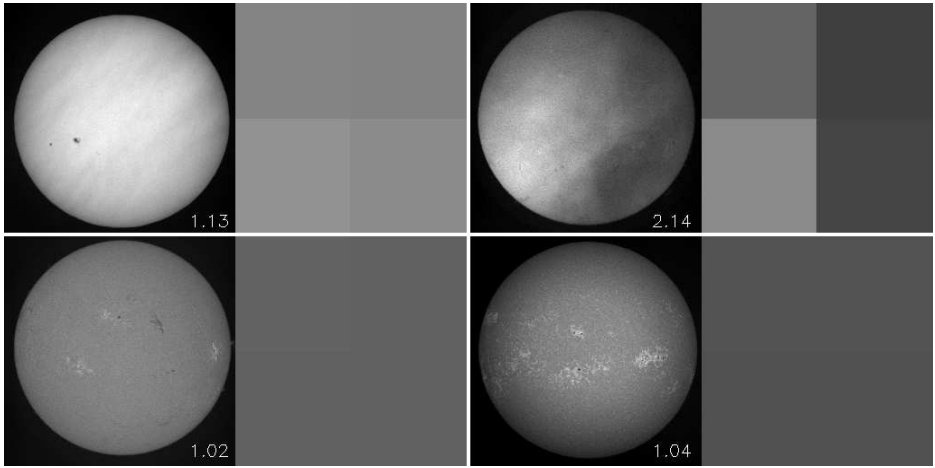}}
 \caption{Four examples for large-scale image inhomogeneity parametrization
 in white-light (upper left, 1~December~2020), \ha\ (upper right --
 31~December~2020, lower left -- 29~January~2016) and \caii\ (lower right --
 7~January~2015) images. The full-disc images are reduced to  2$\times$2 
 pixels (right to each solar disc image) and the ratio of  the maximum 
 to the minimum of the 4 pixels is computed (annotated in each image). 
 The images that are partially obstructed by clouds lead to high ratios,
 far higher than the optimum case of 1.}
 \label{fig:fpr}
\end{figure}

The detection of image inhomogeneities is normally a very computation 
expensive task, which can be done with structural bandpass filters  
({\it cf.} \opencite{Poetzi2015}) or also with neural networks 
(\cite{Jarolim2020}, see also Section \ref{GAN}). For the real-time 
processing this method must be fast and effective. To accomplish this 
task, in the current implementation at KSO,  the recorded and centred image is reduced 
to 2$\times$2 pixels and the ratio of the maximum to the minimum value of these 
pixels, the so called global quadrant intensity ratio ($fpr$), is computed.
Images without any inhomogeneities should have a $fpr$ of nearly 1, active 
regions or filaments have only a small influence on this ratio as their 
area is in the range of maximum a few percent of the solar hemisphere. Clouds, however, can shift 
this ratio to values far larger than 1 as illustrated in Figure \ref{fig:fpr}
top right. A drawback of this method is that clouds uniformly distributed also 
lead to a value of about 1 (Figure \ref{fig:fpr} top left). In such  
case other measures are needed to quantify the reduced image quality 
like the exposure time or the image sharpness.

\subsubsection{Parametrization of image sharpness}\label{detsharp}

Normally for the image sharpness the rms of the intensity distribution
is used, but in case of the Sun this value changes considerably with 
the solar activity. The presence of filaments, active regions or flares 
produce substantially higher rms values compared to a quiet Sun. 
Therefore, for an alternative approach that works fully automatic, we use 
the central part of the solar disc, covering the central quarter of the 
image area, smooth it by convolving the subimage with a simple kernel consisting 
only of 1's, {\it e.g.}, a kernel of size 3 looks like:
\begin{equation}
k_3 = \frac{1}{3^2}
\left[ \begin{array}{ccc}
 1 & 1 & 1 \\ 1 & 1 & 1 \\ 1 & 1 & 1 \\
\end{array}\right]
\end{equation}
For KSO images the kernel $k_{15}$ is used ($\approx$15\,arcsec), which
is the same as averaging pixels in a neighbourhood of 225 pixels. 
This smoothed subimage is then correlated 
with the original subimage. Sharpness values $S$ are then obtained by:
\begin{equation}
S = 1\,000 \times (1 - r)
\end{equation}
where $r$ is the Pearson's correlation coefficient between the smoothed 
and the original subimage. The factor of 1\,000 was introduced to obtain
integer results in a typical range up to $\approx$100.
The reasoning behind this method is the 
following: when the smoothing of the image does not substantially change the image, 
{\it i.e.} when the image was already unsharp before smoothing, 
the value of $r$ would be 1 and therefore $S=0$. Typical values are
$S>50$ in case of quality class 1 \ha\ images, whereas images 
in quality class 3 have $S < 25$. These boundaries are lower in case of 
white-light and \caii\ images.

\subsubsection{Summary of quality determination}\label{sumqual}

The three quality levels are determined by single or combinations of 
the above parameters or procedures, {\it e.g}, for \ha\ images:
\begin{itemize}
  \item[] high exposure times ($exp \ge18$\,ms): quality class 3
  \item[] rms of solar radius determination $>5$ pixel: quality class 3
  \item[] global quadrant intensity ratio $>1.08$: quality class 3
  \item[] $exp$ $>5$\,ms and rms of solar radius $>$ 3: quality class 3 
  \item[] mean intensity in AOI $\le 400$ or $\ge 1200$ counts: quality class 3 
  \item[] image sharpness $<25$: quality class 3
  \item[] image sharpness $<50$: quality class 2
\end{itemize}

The settings for white-light and \caii\ images differ slightly, like the mean 
intensity in the AOI has a much higher default value for white-light images
as we do not expect large changes of image brightness, in contrast to \ha\ 
images that may vary strongly during flares. 
The parameters defining the
quality classes for each camera are listed in Appendix Section \ref{qualparam}.

\subsection{Quality parametrization with Generative Adversarial Networks}\label{GAN}

\begin{figure} 
 \centerline{\includegraphics[width=1.0\textwidth,clip=]{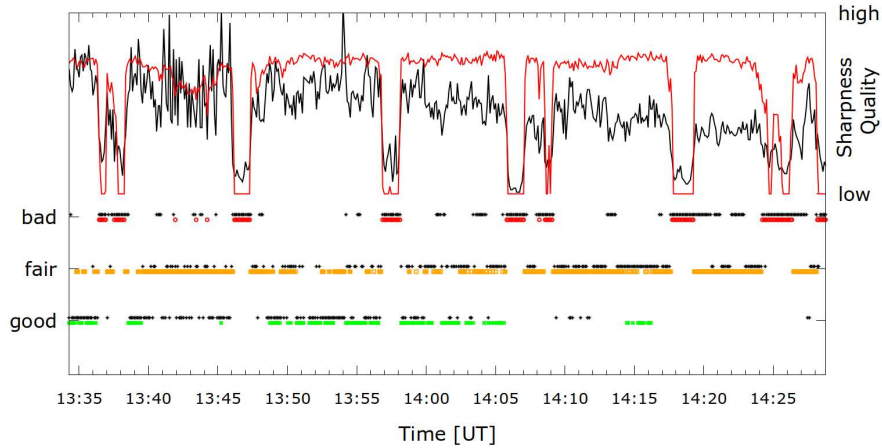}}
 \caption{Comparison of two different quality parametrization methods
 applied to KSO \ha\  images on 26~January~2019.
 In the upper panel the quality value obtained with the GAN method (red) 
 is compared to the sharpness value $S$ (black). The lower panel shows 
 that the quality estimations dividing the images into three classes 
 of both methods are in general in agreement. }
 \label{fig:qualcomp}
\end{figure}

A different approach for image quality assessment of ground-based 
solar observations was presented by \cite{Jarolim2020}, where the 
authors used a neural network to identify quality degradation. 
The method is based on generative-adversarial-networks (GANs) to model 
the true image distribution of high quality KSO \ha\ observations and 
assessing deviations from it, such that atmospheric influences 
({\it e.g.}, clouds) result in large deviations. This metric is 
more objective in terms of the solar activity level. The method 
provides a human-like quality assessment that agrees in 98.5\% of 
the cases with human assigned labels. 
The developed neural network provides a continuous image quality 
metric and a binary classification into high- and low-quality 
observations. In the upper panel of Figure \ref{fig:qualcomp} we 
compare the image sharpness (defined in Section \ref{detsharp}) with the 
continuous image quality value of the neural network. For comparison 
fixed thresholds in the image quality metric were set and additionally
the binary classification to further distinguish between good (class 1) 
and fair (class 2) images was used ({\it cf.} \opencite{Jarolim2020}). 
The direct comparison shows 
that both assessment methods are in general in good 
agreement. The neural network shows a higher sensitivity for clouds, 
while it is less sensitive to the variations in seeing, in agreement 
with the objective for large scale image inhomogeneities. The fast 
performance of the neural network makes it suitable for the critical 
step of quality assessment in a real-time environment. The further 
extension to small scale quality estimations ({\it e.g.}, seeing), 
multiple channels ({\it e.g.}, white-light, \caii) 
and the applicability to a real-time data pipeline are under investigation.

\subsection{Correction of large-scale image inhomogeneities}\label{inhomoCorr}

\subsubsection{Centre-to-limb variation}

In order to derive a centre-to-limb variation (CLV) profile we apply 
a median filter on each concentric ring of the solar disc with a width 
of one pixel. Features like sunspots, active regions or filaments 
do in general not have an effect on such a filter, as they are clearly smaller 
than the half length of such a concentric ring. However, close to the 
centre of the solar disc, these activity features can cover a substantial 
part of the rings. Thus, for robustness, for the inner rings a 
further post-processing should be applied, {\it e.g.}, an interpolation via a 
CLV functional fitting as is described below. 
This procedure is computationally intensive, but as each ring can 
be handled independently the code can easily be parallelized.

According to \cite{scheffler1990} the CLV profile can theoretically be 
expressed as 
\begin{equation}
I_{\beta} = \frac{1+\beta\cdot\cos \theta}{1+\beta}
\end{equation}
where $I$ is the intensity, $\beta$ is the  centre-to-limb variation factor 
and $\theta$ is the exit angle of the light (disc centre: $\beta = 0$, 
limb: $\beta =\pi/2$). 
In order to use distances from the centre 
of the solar disc this equation can also be rewritten to
\begin{equation}\label{Eq:clv}
I_x = \frac{1+b\cdot\cos(\arcsin(a\cdot x))}{1+b}
\end{equation}
with $a$ as scale factor of the solar radius, $b$ as centre-to-limb 
variation factor and $x = r/R_{\odot}$. 
This formula does not describe the whole profile as the 
numerator is always larger than or equal one and the upper limit of the 
denominator is defined by $1+b$, which is in most cases smaller than 3.

\begin{figure} 
\centering
  \includegraphics[width=\textwidth,clip=]{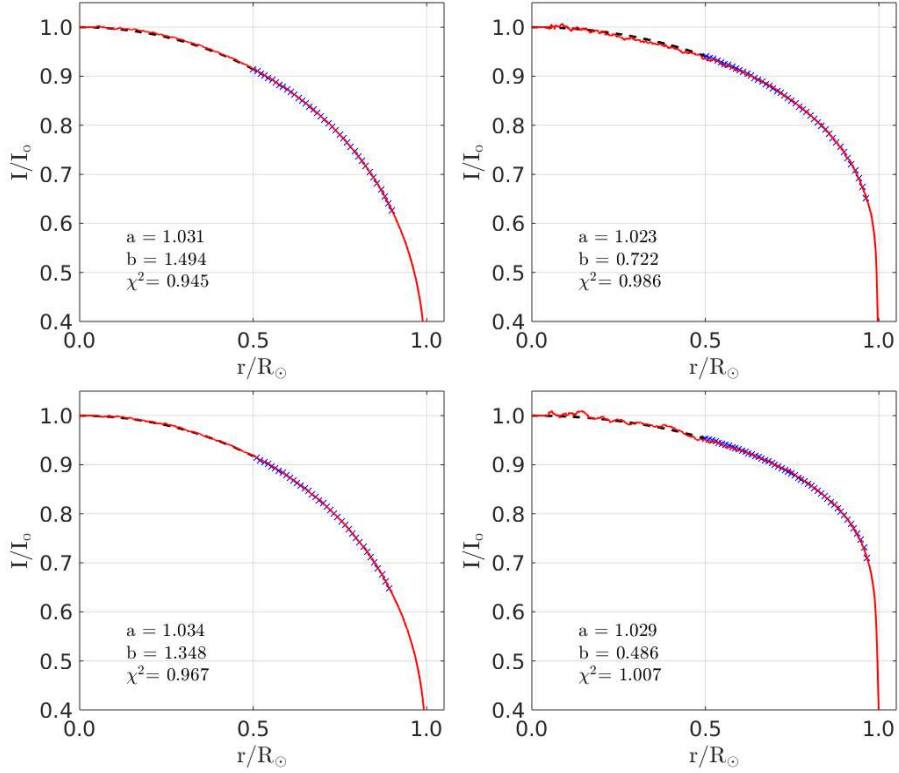}
 \caption{Radial CLV intensity profile derived from KSO white-light 
  (left) and \ha\ (right) images. In the upper panel the images were 
  selected during quiet Sun conditions on 1 August 2017. The lower 
  panel shows the profiles for 22 October 2014, when a very large
  sunspot group was present. The red curve is the CLV profile calculated 
  from the median intensity of the rings around Sun centre,
  the blue crosses correspond to the CLV functional fit to the data 
  according to Eq. \ref{Eq:clv} for $r > 0.5R_{\odot}$, and the 
  dashed line is the extrapolation of the CLV fit to $r < 0.5R_{\odot}$.}
 \label{fig:clvwlhalph}
\end{figure}

\begin{figure} 
\centering
  \includegraphics[width=\textwidth,clip=]{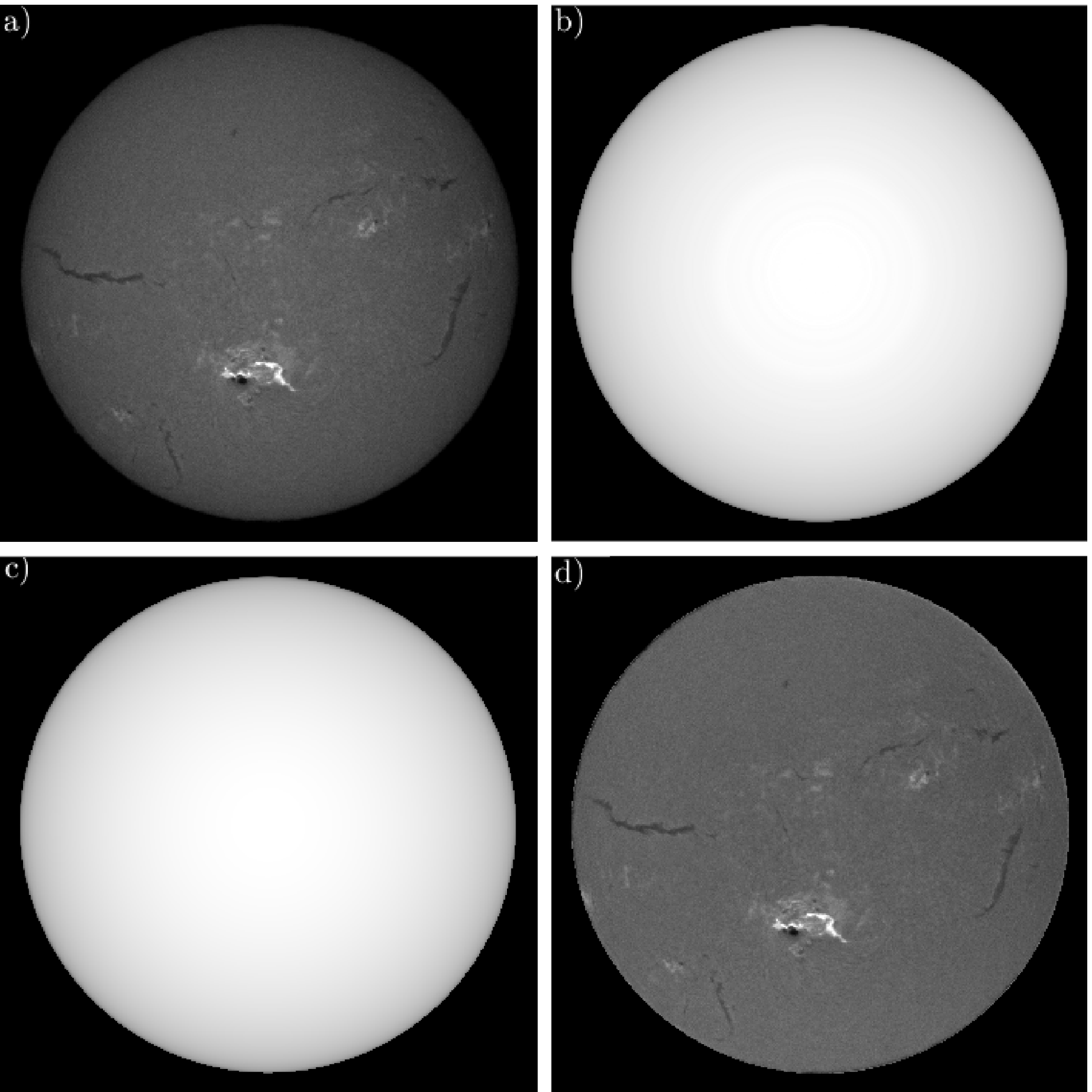}
 \caption{KSO \ha\ flare image on 22~October~2014 at 14:55:05 UT. 
 a) Original image,  b) raw CLV determined by the median intensity along 
 rings around the solar disc centre, c) CLV$_{new}$ obtained by the 
 functional CLV fit (Eq. \ref{Eq:clv}) and extrapolation for $r>0.5 R_{\odot}$, 
 and d) image divided by 
 CLV$_{new}$. In the raw CLV map the intensity variations between the 
 individual rings are visible, these are smoothed out in CLV$_{new}$ 
 by applying the functional fit. The contrast in image d) is
 constant from the disc centre to the limb.}
 \label{fig:correctclv}
\end{figure}

By fitting the calulated CLV profile by Eq. \ref{Eq:clv}, a complete
smooth CLV profile can be obtained which is not affected by intensity fluctuations
introduced by the individual rings. In order to avoid the potential effects 
of solar activity features (sunspots, flares, filaments) on the inner 
rings, we applied the fits to different subranges  (avoiding the 
inner parts) and evaluated the robustness of the fit results, 
namely  $[0-1]R_{\odot}$, $[0-0.97]R_{\odot}$, $[0.10-0.97]R_{\odot}$, 
$\ldots [0.70-0.97]R_{\odot}$. 
These evaluations were done with KSO white light and \ha\ mages recorded 
at times of very low activity (no sunspots present) and the resulting 
fit range was then also used for images obtained at times of high activity.
The fit result is extrapolated on the full range using Eq. \ref{Eq:clv}
giving the new centre-to-limb variation (CLV$_{new}$) profile. 
This method also works for active solar conditions, {\it i.e.} images with 
magnetic features like sunspots in white-light or active regions in \ha. 
An example for this procedure for very high solar activity with one of 
the largest sunspot groups in the last decade on 22 October 2014 and for 
a quiet Sun from 1 August 2017 is show in Figure \ref{fig:clvwlhalph}.
This approach assures that large sunspots, filaments and flares close 
to the inner (smaller) rings, where they may produce a substantial effect 
on the obtained (median) ring intensities, do not affect the determined 
CLV profile. The application of this procedure to produce a smooth CLV is 
demonstrated in Figure \ref{fig:correctclv}. One can see that in the inner 
regions, the raw CLV profile calculated from the median intensities 
along the rings rings is affected by the flare (panel b). This effect 
is removed by the CLV fit and extrapolation for the inner parts $r>0.5R_{\odot}$ 
(panel c). Also, the image contrast is improved considerably by dividing 
the original image with the CLV$_{new}$ intensity profile (panel d). We note that 
such strong deviations show up in the presence of large flares in \ha\ 
images, due to the large intensity enhancement during flares. Big 
sunspots in white light images do in general not reveal a significant distortion 
of the inner CLV rings.

\begin{figure} 
\centering
  \includegraphics[width=0.7\textwidth,clip=]{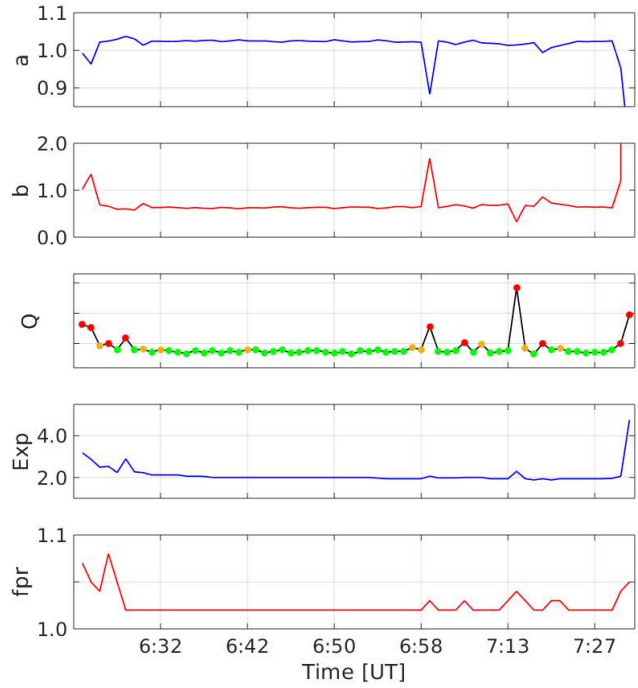}
 \caption{Evolution of KSO image parameters during one hour
 of \ha\ observations  on 19~May~2020. From top to bottom: CLV parameters $a$ and $b$, 
 image quality parameter from GAN ($Q$), exposure time ($exp$) and 
 global quadrant intensity ratio ($fpr$). Deviations of the parameter 
 $a$ and $b$ are linked to lower image quality $Q$ ({\it i.e.}, green
 dots for quality class 1 and orange or red dots for classes 2 and 3, respectively).}
 \label{fig:parquality}
\end{figure}

Finally, the CLV profile fit parameters $a$ and $b$ were 
also studied in their evolution over complete observing days and 
compared to different other image parameters, such as exposure time ($exp$), 
the global image quadrant intensity ratio ($fpr$), the image sharpness parameter
and the image quality ($Q$) from \cite{Jarolim2020}. Figure \ref{fig:parquality} 
shows as an example comparison of these parameters over about one hour 
of \ha\ observations. As can be seen, the CLV parameters evolution during the day also 
delivers an assessment of the image quality. In general, values 
deviating from average $a$ and $b$ are related to lower image quality.

\subsubsection{Global anisotropies}

\begin{figure} 
 \centerline{\includegraphics[width=0.5\textwidth,clip=]{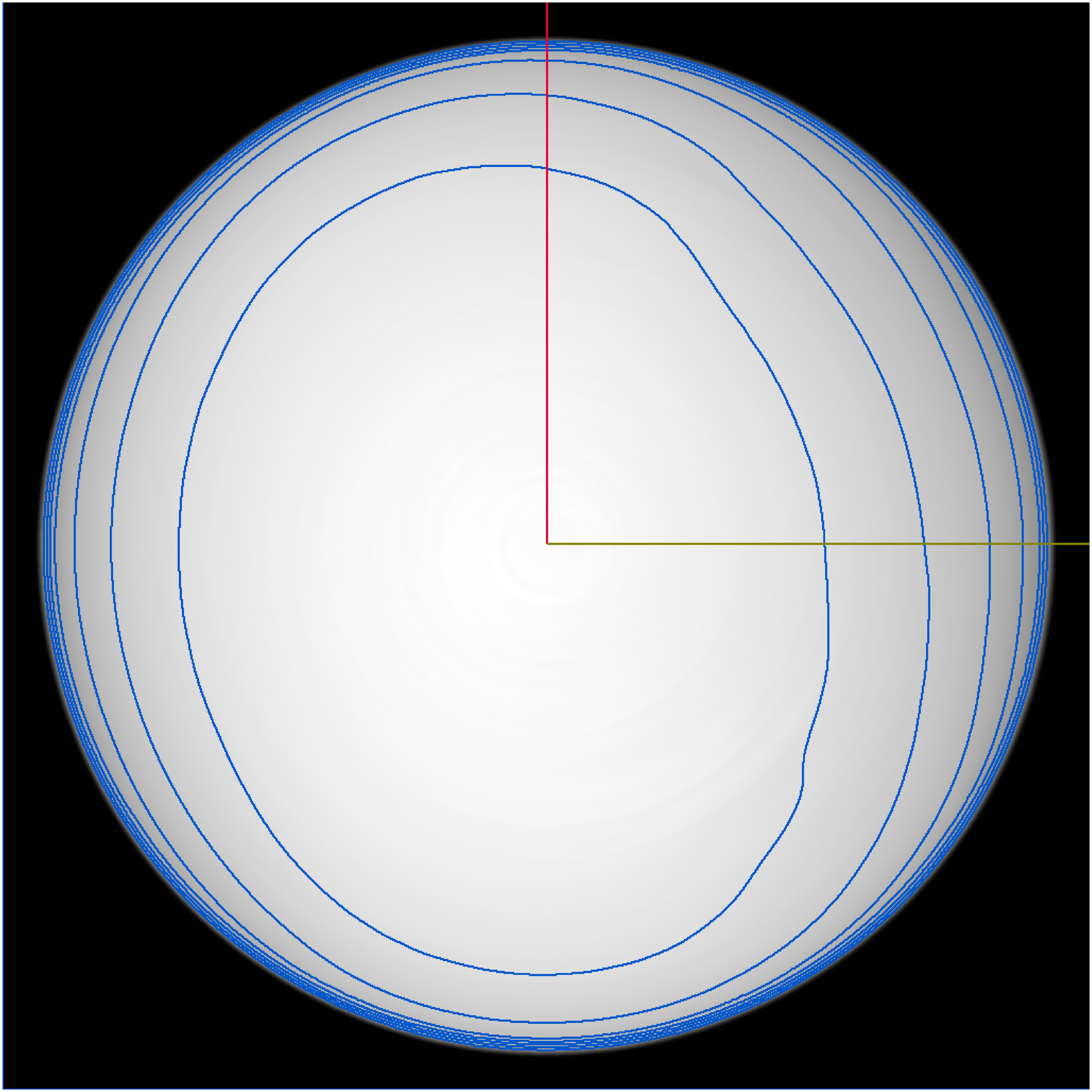}
 \includegraphics[width=0.5\textwidth,clip=]{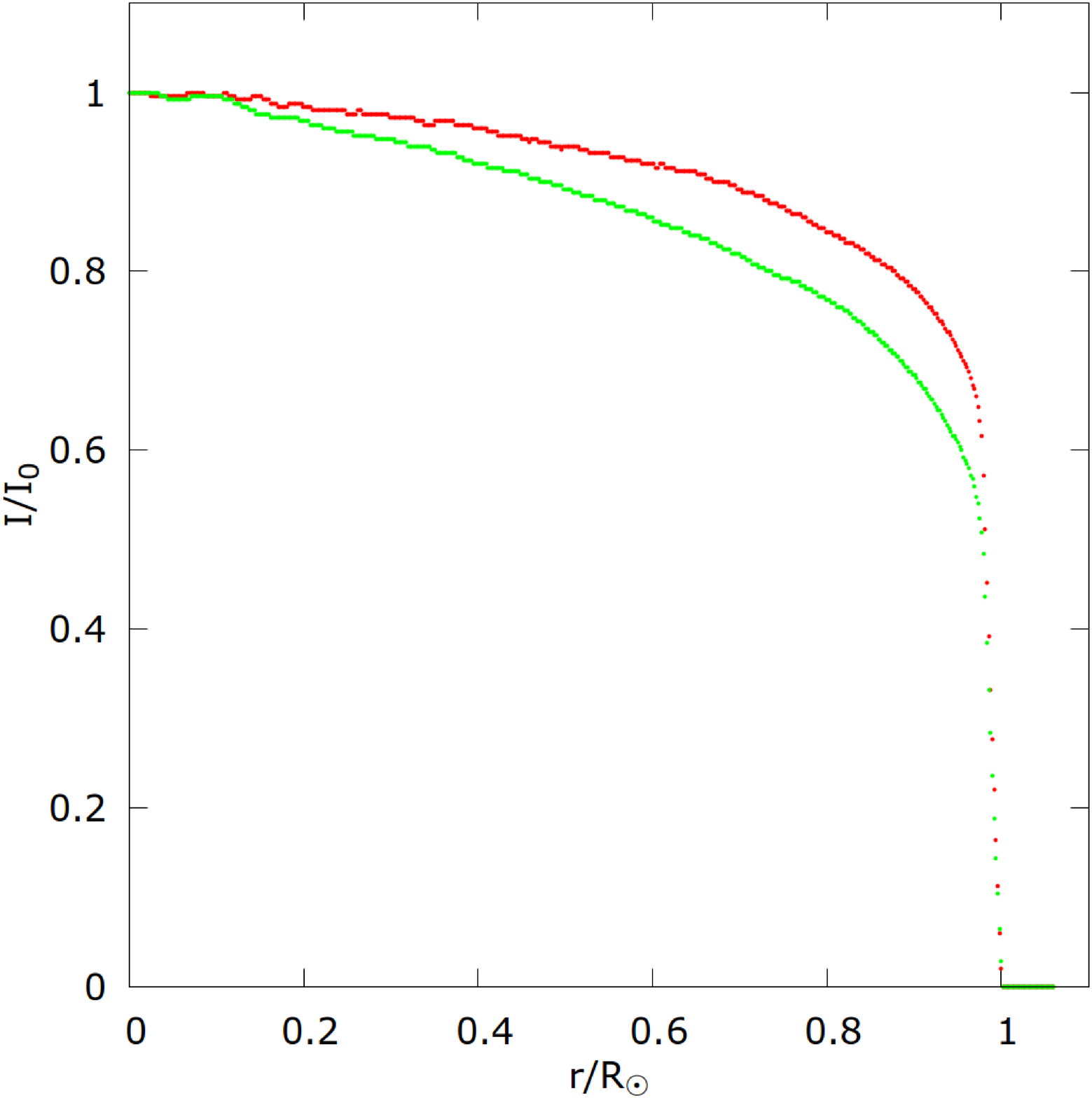}}
 \caption{Global anisotropies map derived from an \ha\ image of 26 January 2021 and the 
 corresponding normalized radial intensity profiles. Due to the optical parts 
 in the telescope the intensity distribution in the map is not radially 
 symmetric, which can be clearly seen in the overlaid contour lines (blue).
 The red and green lines correspond to the intensity profiles shown in the 
 right panel.}
 \label{fig:aniso}
\end{figure}

Normally the telescope optics introduces additional intensity inhomogeneities.
This can be due to dust, stray-light, filters or even shutters (a good 
example can be seen in \citet{Denker1999}). 
In order to remove also such influences the method of obtaining the
CLV profile has to be adapted by replacing the median of each 
concentric ring with a running median. This method was first used by
K.\,A. Burlov-Vasiljev at Kiepenheuer Institute in Freiburg, Germany.
For this purpose the filter width of the running median filter 
applied to each concentric ring has to be set in order to neglect solar 
features but still detect large scale inhomogeneities.
A filter width of twice the radius of each concentric ring  ($\approx$ one
third of the length of the ring) 
seems to be a choice that fullfills these criteria. The central part 
($r<R_{\odot}/20$) is treated as for the CLV determination, where the median
of the whole disc inside this boundary is taken. In the \ha\ images at KSO 
there is a small intensity gradient from east to west (left to right), 
this can be seen in the contour lines of the anisotropy map of Figure 
\ref{fig:aniso}. The right panel illustrates this gradient more apparently.
The red profile from centre to north looks like a standard CLV profile, 
whereas the green profile from centre to west drops down too rapidly. 

\begin{figure} 
 \centerline{\includegraphics[width=0.5\textwidth,clip=]{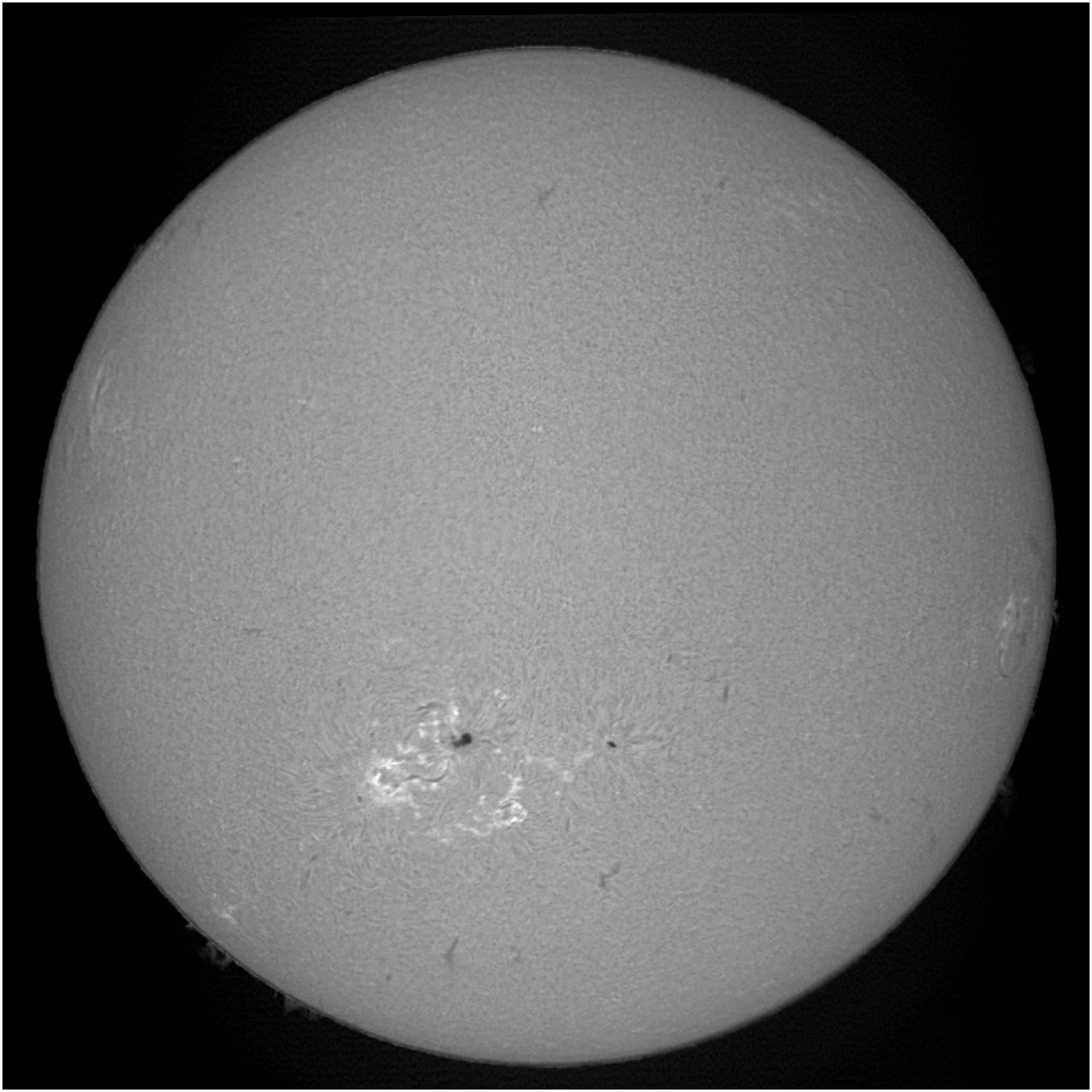}
 \includegraphics[width=0.5\textwidth,clip=]{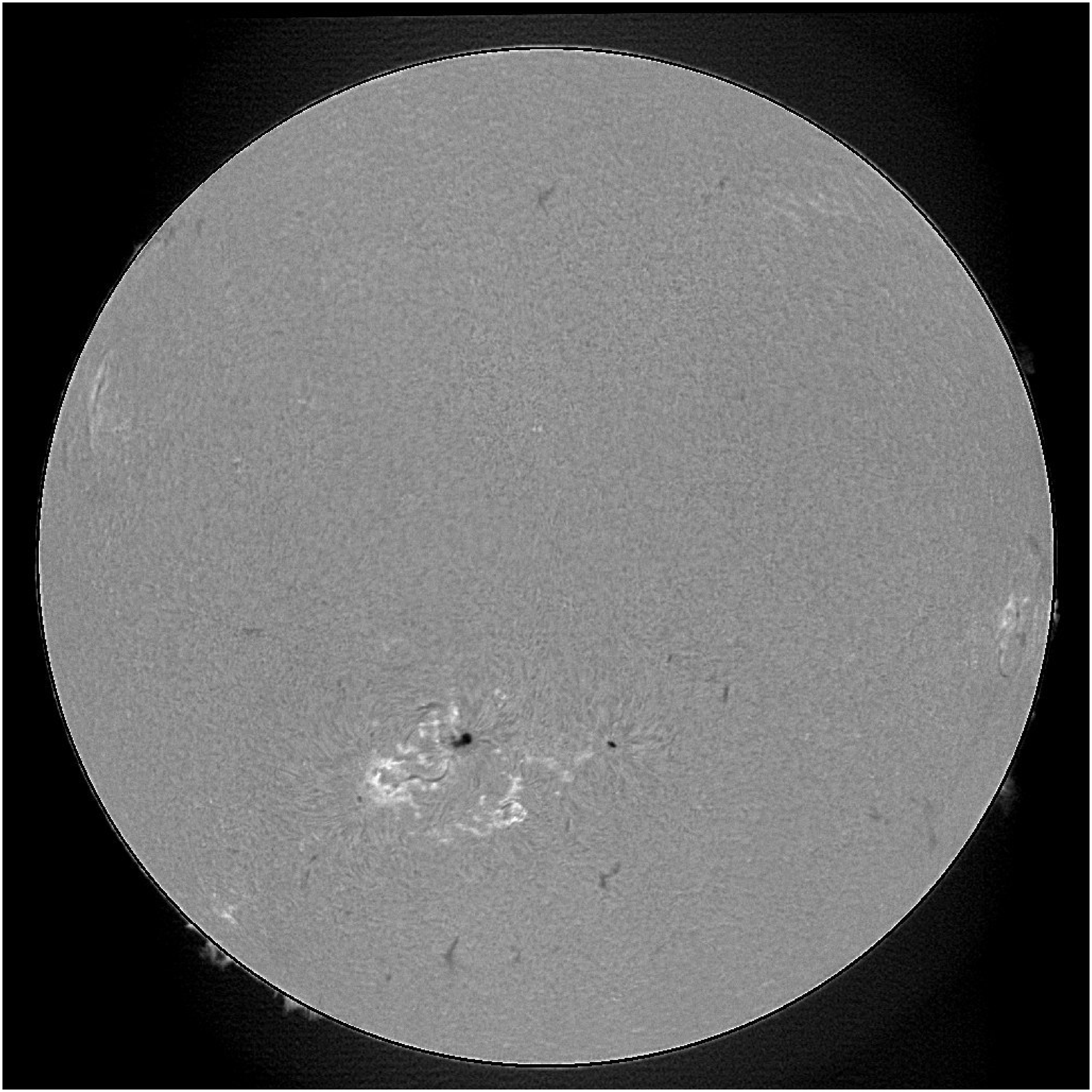}}
 \centerline{\includegraphics[width=0.5\textwidth,clip=]{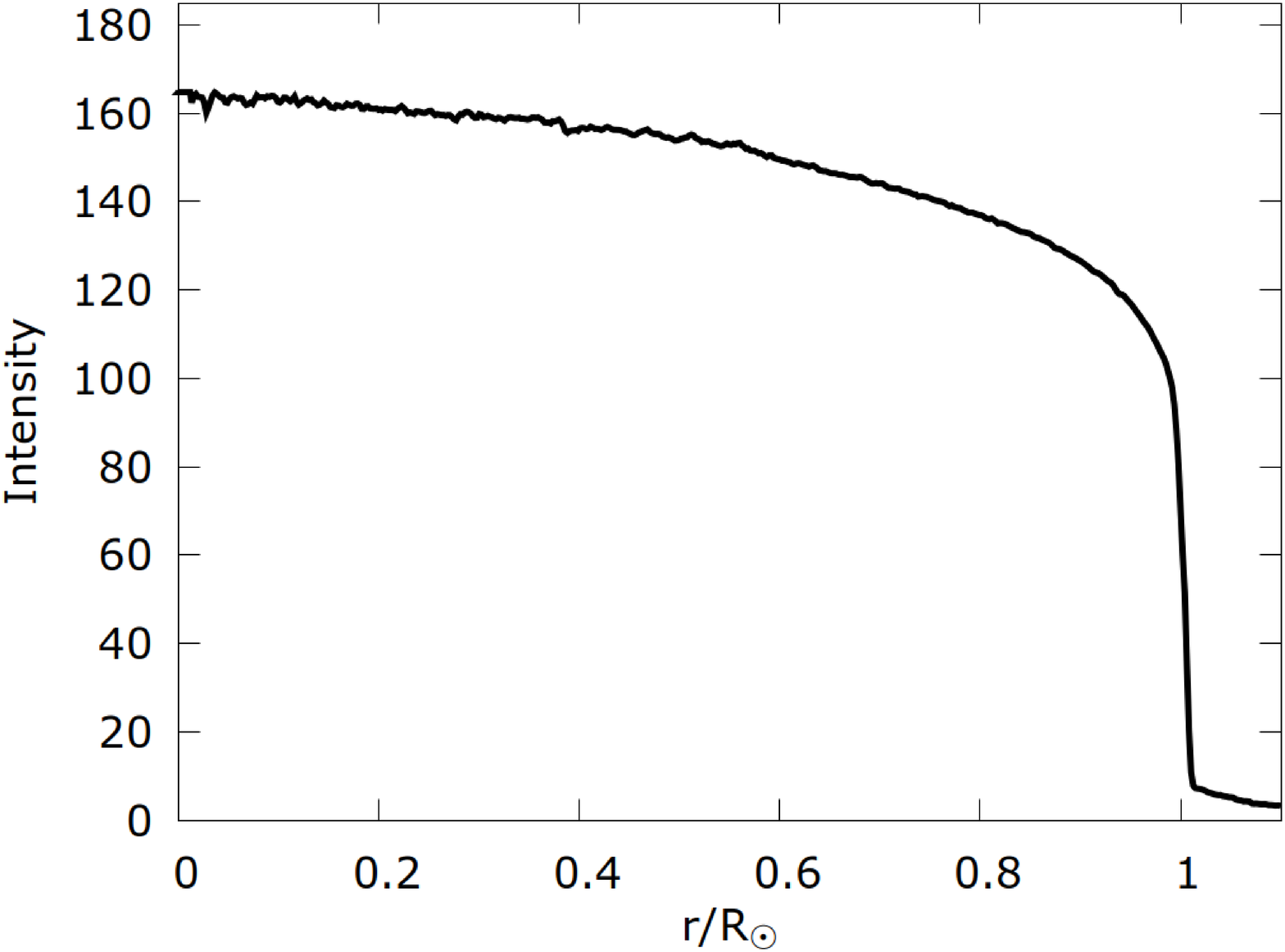}
 \includegraphics[width=0.5\textwidth,clip=]{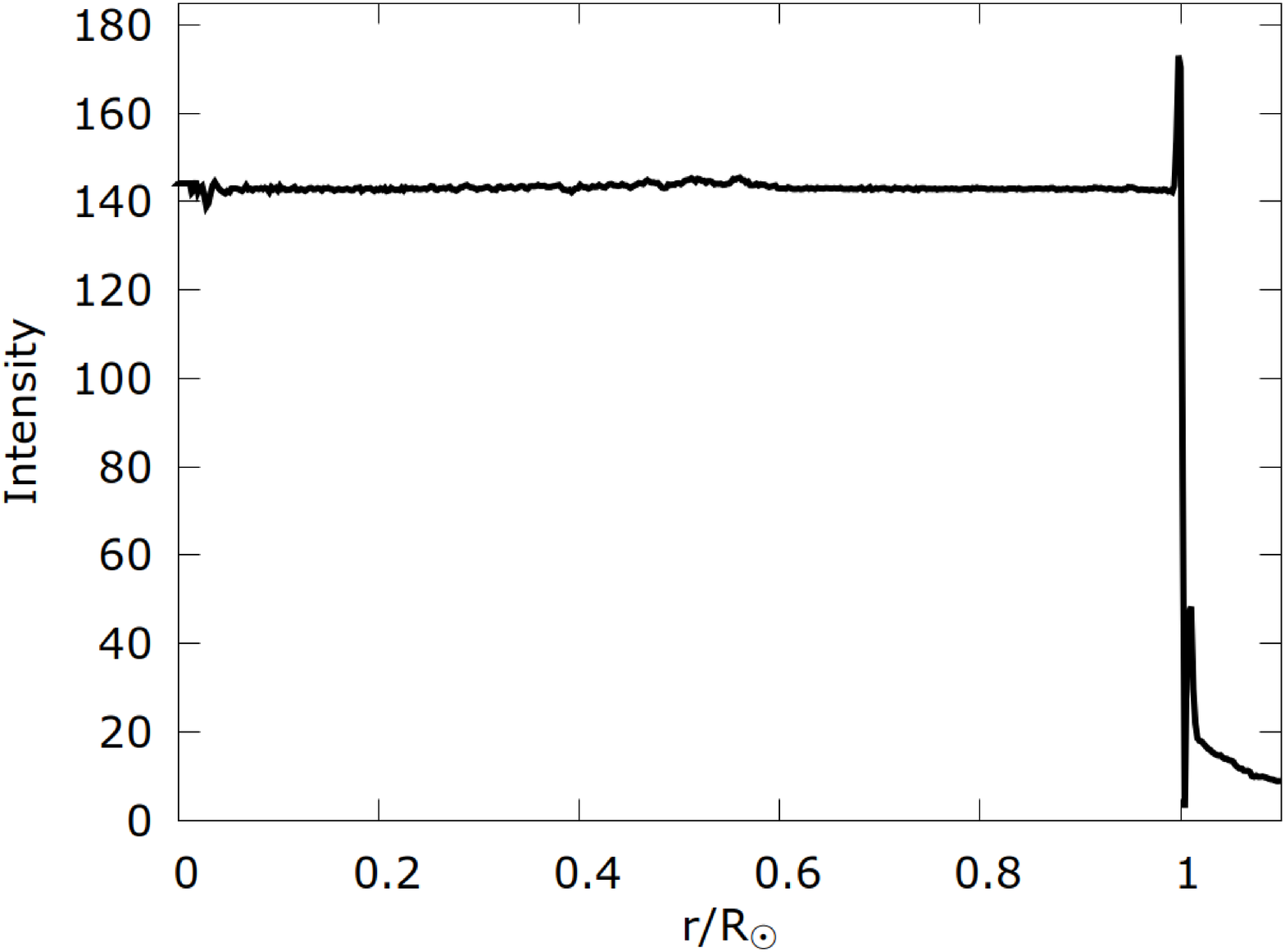}} 
 \caption{\ha\ images for comparison between normal contrast image (left)
 and high contrast image (right) divided by an anisotropy map
 from 28~November~2020.
 The radial intensity profile under each image shows that the corrected image
 is nearly flat after the correction. The wiggly structure near $r/R_{\odot}\approx0$ 
 is due to the little number of averaged pixels. The jump at the limb in the 
 right profile is a result of the extra treatment of the scaling inside the 
 solar disc and above the solar limb.}
 \label{fig:contr}
\end{figure}

To get a widely flat image -- having a uniformly distributed intensity
level and everywhere almost the same image contrast -- the image has to 
be divided by the obtained anisotropic map. Figure \ref{fig:contr} shows 
a comparison between the original image (left) and the flat image (right).
Especially features at the limb, like the small active region in the
west (right), become better visible. The radial intensity profiles shown 
beneath the image demonstrates the flatness of the image,
the higher values and the variations in the central part ($r/R_{\odot}\approx0$) 
are a result of the small number of averaged pixels. The variations near
$r/R_{\odot}=0.55$ are introduced by the large active region. The peak at the 
limb of the corrected profile is due to the roughness of the solar limb caused 
by seeing effects. The jump at $r/R_{\odot}>1.0$ in the intensity profile 
is a result of the scaling above the solar limb in order to enhance prominences.

\subsubsection{Flat field determination}\label{subFF}

The Kuhn-Lin-Loranz \citep{Kuhn1991} method is used to obtain a real flat 
field, that also includes smaller intensity variations, like dust particles: 
This method is based on the assumption that the Sun does not change its 
appearance within a few seconds. A set of Sun images shifted on the CCD is used to 
calculate the intensity for each CCD pixel. The more images are used and 
the less the Sun changes, good seeing conditions are a premise, the better 
the flat field becomes. A drawback of this method is, that this procedure needs 
manual interaction, and cannot be performed very often as it interrupts 
the patrol observations.

\begin{figure} 
 \centerline{\includegraphics[width=1\textwidth,clip=]{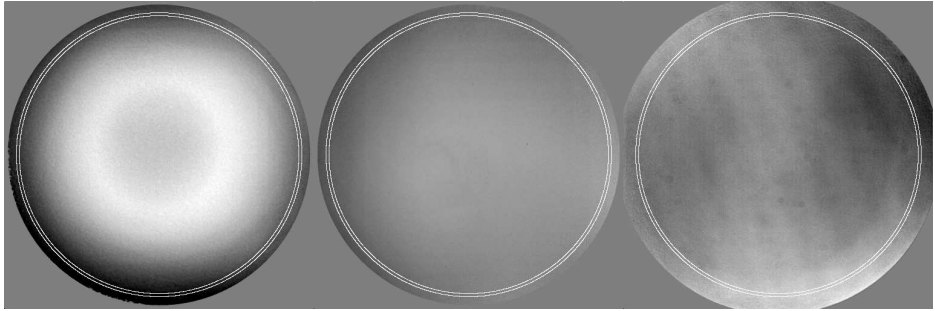}}
 \caption{Flat field frames for \caii\ (left, 9~April~2021), 
 white-light (centre, 9~May~2012) and \ha\ (right, 3~June~2021) telescopes.
 The contrast is strongly enhanced to see the inhomogeneities better. 
 The white rings represent the limits of the solar disc size in  aphelion and perihelion.
 Missing values and regions far above the solar limb are set to 1.
 The enhanced intensity in a concentric ring at about $R_{\odot}/3$ in the
 \caii\ flat field is due to  the filter. The interference pattern in \ha\ is
 most likely produced within the Lyot filter.}
 \label{fig:ff}
\end{figure}

At KSO for such a flat field one centre image plus 28 shifted images are 
taken. The shifted images are on three concentric rings, four images with
a  shift of 115 pixel ($\approx R_{\odot}/8$), 8 images with a shift of 
230 pixel and 16 images with a shift of 345 pixel. The exposure time is set to 
a fixed value and the duration of the whole procedure is kept as short as 
possible in order to have almost the ''same'' Sun for all images. Taking 
the full frameset takes about four minutes, most of the time is used for 
moving the telescope.  Starting with 2021, flat field frames are produced 
for all telescopes (\ha, \caii\ and white-light), when the seeing 
conditions are exceptionally good or when there are any changes in the 
telescope setup.  Figure \ref{fig:ff} shows an example
for flat field frames of all three telescopes.
For \caii\ images the interpolation of the CLV according to Eq. \ref{Eq:clv}
does not work properly. The reason for this is the intensity distribution 
of the flat field, which seems to be produced mainly by the 
\caii\ filter. This flat field shows a ring-like intensity enhancement at 
$R_{\odot}/3$, the central intensity is even lower than in this ring. 
The white-light flat field (centre panel of Figure \ref{fig:ff}) shows a
relatively uniformly distribution, some very small dust particles are visible.
The \ha\ flat field (right panel of Figure \ref{fig:ff}) is dominated by three 
features: The interference pattern (dark and bright bands) are very likely 
produced by the Lyot filter, additionally there is a global intensity gradient
from the centre to the left and to the right and there are some dust particles.
We have to note that the contrast in these flat field frames was strongly 
enhanced to make all these features visible, the intensity difference within
the solar disc is about two percent.
The horizontal and vertical structures are an artefact due to the image motion
caused by atmospheric conditions, which should be theoretically zero for 
applying the Kuhn-Lin-Loranz method. Also the small scale structures are a 
result of the seeing. To get rid of these artefacts a really exceptionally 
good seeing is necessary for getting a good flat field with this method. 
The most important application for such a flat field map is to detect dust 
particles on the optical parts and to see the influences of the telescope
optics on the camera image. Applying such a flat field on the raw image can
only be done after some smoothing steps in order to get rid of the fine
structures introduced by the seeing.


\section{Data Provision}

\subsection{KSO data archive}\label{subArch}

The image archive of KSO as presented in \cite{Poetzi2013} underwent a
number of changes and updates in the recent years. The most important 
improvement is the instant of time of availability of the data in the archive. 
Whereas the archive was normally updated at the end of an observing day, 
it is now updated continuously in quasi real-time simultaneously to 
the observations. Every minute, for each telescope one newly 
recorded image is transferred to the archive. As there are differences
in the image cadence (10 images per minute for \ha\ and \caii\ and 3 images per minute for
white-light), additional recorded images are transferred to the archive after
the observing day. The number of additional images to be transferred is 
selected depending on camera type and solar activity. Normally three 
chromospheric (\ha\ and \caii) images per minute are stored and one
photospheric (white-light). In case of large solar flares\footnote{
Selected from NOAA (National Oceanic and Atmospheric Administration/Space 
Weather Prediction Center) solar and geophysical event reports: 
\url{https://www.swpc.noaa.gov/products/solar-and-geophysical-event-reports}} 
(defined of GOES class $\ge$ M1 or optical flare importance class $\ge 1$) 
all images of good and fair quality recorded during flare activity are 
transferred to the archive.

\begin{figure} 
 \centerline{\includegraphics[width=1.0\textwidth,clip=]{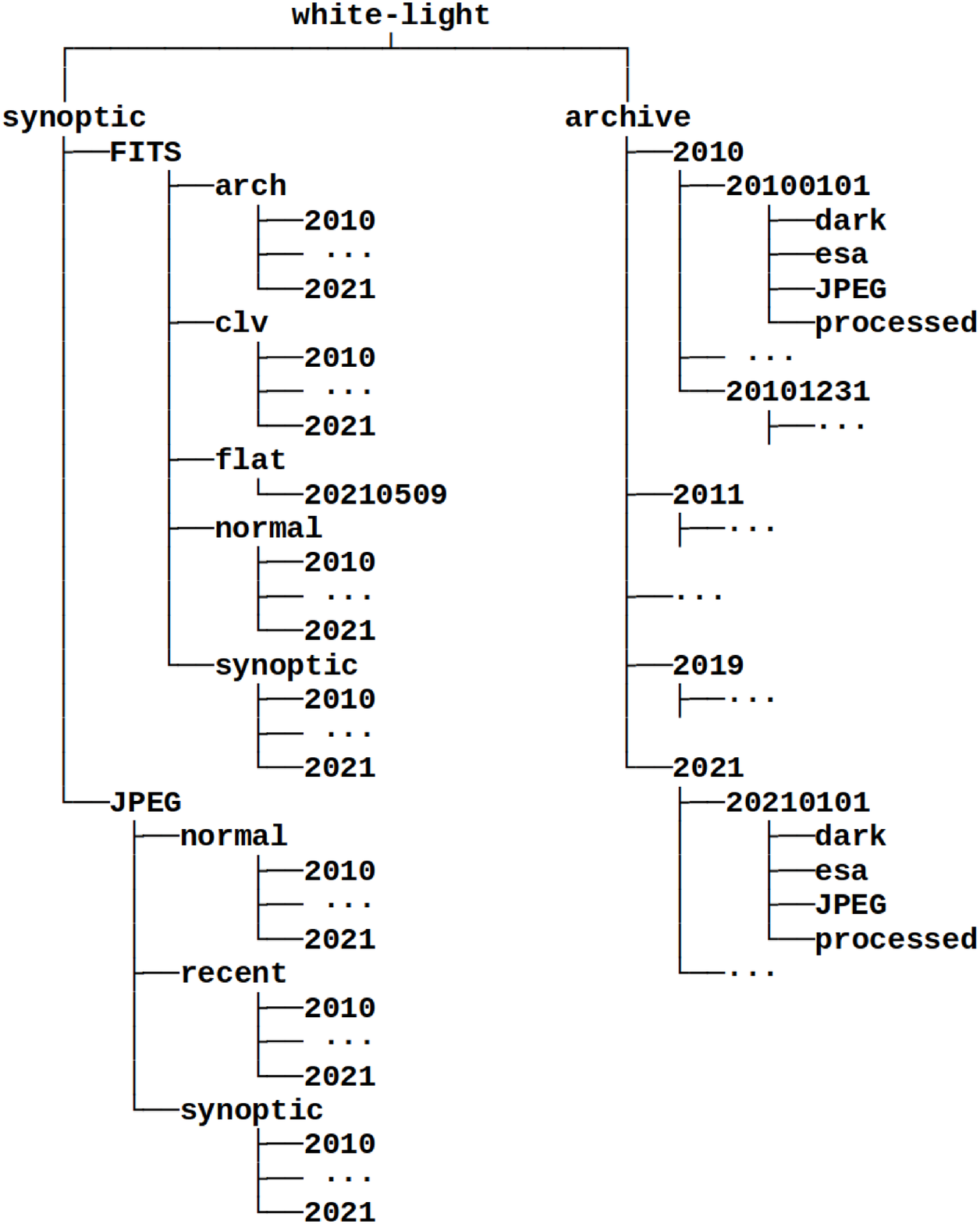}}
 \caption{The data path for the full white-light data set. A data set of 
 one daily processed image is in the {\sl synoptic} path, the whole
 daily data is located in the {\sl archive} path.}
 \label{fig:paths}
\end{figure}

\begin{figure} 
 \centerline{\includegraphics[width=1.0\textwidth,clip=]{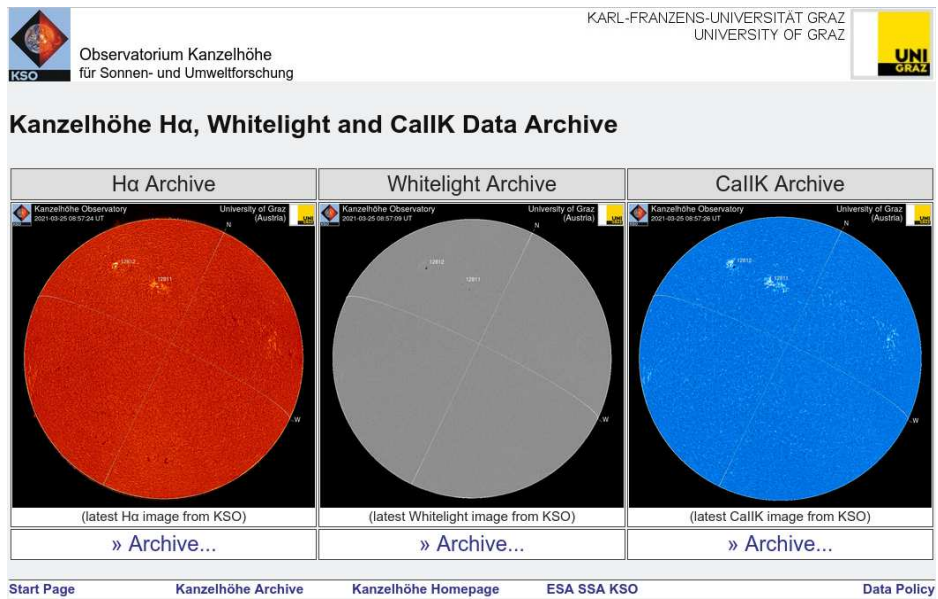}}
 \caption{The mirror archive located physically at University of Graz 
 is operated for fast data access. In case of large data transfers this 
 archive should be favoured.}
 \label{fig:grazmirr}
\end{figure}

The data archive consists of the synoptic and the main archive 
branch (see Figure\ref{fig:paths}). The synoptic archive consists of one 
higher level image data set for each observing day. The main archive 
holds all the observed raw images as FITS files and their corresponding 
processed JPEG quick look files. The data hierarchy in the synoptic branch 
goes from data type to yearly directories. In the main archive the data 
are stored in yearly directories containing the folders for 
each observing day. All archive data are mirrored in a local archive 
simultaneously. With a delay of a few hours, the data are also 
transferred to a mirror archive located physically at the University 
of Graz (Figure \ref{fig:grazmirr}), which provides a higher bandwidth
for faster data transfer. The archive can be accessed via following ways:
\begin{itemize}
  \item KSO archive webpage \url{http://cesar.kso.ac.at}
  \item KSO ftp \url{ftp://ftp.kso.ac.at}, instructions how to access the server 
  can be found on  \url{http://cesar.kso.ac.at/main/ftp.php}
  \item for the transfer of large data sets we suggest the use of the mirror 
    \url{http://kanzelhohe.uni-graz.at} (Figure \ref{fig:grazmirr}) as
    it guarantees faster download rates.
  \item ESA SSA Spaceweather portal \url{https://swe.ssa.esa.int},
   registration is necessary to access this data.
\end{itemize}

\subsubsection{Synoptic archive data products}

The directory structure for the synoptic archive shown in Figure \ref{fig:paths} 
(left) consists of a FITS and JPEG path for each camera containing 
a set of files for each observing day. The image is selected manually by
the observer and set of various images and profiles is processed for displaying 
on the observatory website. These data are for a quick daily overview 
of the situation on the Sun.

\begin{itemize}
    \item[The following files are generated:] 
    \item[]
    \item {\sf kanz\_[type]\_fi\_[yyyymmdd]\_[HHMMSS].fts.gz}
      \begin{itemize}
        \item the raw selected image with updated FITS header
        \item located in {\sf FITS/arch/[yyyy]}
        \item example:\href{http://cesar.kso.ac.at/caiis/FITS/arch/2021/kanz_caiik_fi_20210425_060742.fts.gz}{kanz\_caiik\_fi\_20210425\_060742.fts.gz}
      \end{itemize}
    \item {\sf kanz\_[type]\_fq\_[yyyymmdd]\_[HHMMSS].fts.gz} 
      \begin{itemize}
        \item the CLV intensity profile
        \item located in {\sf FITS/clv/[yyyy]}
        \item example:\href{http://cesar.kso.ac.at/halpha3s/FITS/clv/2015/kanz_halph_fq_20150107_120351.fts.gz}{kanz\_halph\_fq\_20150107\_120351.fts.gz}
      \end{itemize}
    \item {\sf kanz\_[type]\_fc\_[yyyymmdd]\_[HHMMSS].fts.gz} 
      \begin{itemize}
        \item a high contrast image {\it i.e.} global anisotropies 
           are corrected and the solar disc is centred.
        \item located in {\sf FITS/normal/[yyyy]} 
        \item example:\href{http://cesar.kso.ac.at/caiis/FITS/normal/2021/kanz_caiik_fc_20210425_060742.fts.gz}{kanz\_caiik\_fc\_20210425\_060742.fts.gz}
      \end{itemize}
    \item {\sf kanz\_[type]\_ff\_[yyyymmdd].fts.gz} 
      \begin{itemize}
        \item the flat field map normalized to 1.0 and the individual 
          raw files used to generate this flat field map.
        \item located in {\sf FITS/flat/[yyyymmdd]}
        \item example:\href{http://cesar.kso.ac.at/caiis/FITS/flat/20210409/kanz_caiik_ff_20210409.fts.gz}{kanz\_caiik\_ff\_20210409.fts.gz}
      \end{itemize}
    \item {\sf kanz\_[type]\_fd\_[yyyymmdd]\_[HHMMSS].fts.gz}
      \begin{itemize}
        \item a normal contrast image with centred solar disc.
        \item located in {\sf FITS/synoptic/[yyyy]}
        \item example:\href{http://cesar.kso.ac.at/phokads/FITS/synoptic/2015/kanz_bband_fd_20151227_0847.fts.gz}{kanz\_bband\_fd\_20151227\_0847.fts.gz}
      \end{itemize}
    \item[]
    \item {\sf kanz\_[type]\_fc\_[yyyymmdd]\_[HHMMSS].jpg}
      \begin{itemize}
        \item a high contrast image {\it i.e.} global anisotropies are 
         corrected, the solar disc is centred, a heliographic grid is 
         overlaid, the image is rotated to North up and annotated.
        \item located in {\sf JPEG/normal/[yyyy]}
        \item example:\href{http://cesar.kso.ac.at/halpha3s/JPEG/normal/2021/kanz_halph_fc_20210425_060748.jpg}{kanz\_halph\_fc\_20210425\_060748.jpg}
      \end{itemize}
    \item {\sf kanz\_[type]\_fc\_[yyyymmdd]\_[HHMMSS].icon.jpg}
      \begin{itemize}
        \item an icon of the high contrast image.
        \item located in {\sf JPEG/recent/[yyyy]}
        \item example: \href{http://cesar.kso.ac.at/halpha3s/JPEG/recent/2021/kanz_halph_fc_20210425_060748.icon.jpg}{kanz\_halph\_fi\_20210425\_060748.icon.jpg}
      \end{itemize}
    \item {\sf kanz\_[type]\_fd\_[yyyymmdd]\_[HHMMSS].jpg}
      \begin{itemize}
        \item a normal contrast image {\it i.e.} the solar 
         disc is centred, an unsharp masking is applied for little contrast
         enhancement, the image is rotated to North up and annotated.
        \item located in {\sf JPEG/synoptic/[yyyy]}
        \item example: \href{http://cesar.kso.ac.at/phokads/JPEG/synoptic/2015/kanz_bband_fd_20151227_0847.jpg}{kanz\_bband\_fd\_20151227\_0847.jpg}
      \end{itemize}
\end{itemize}
Where [type] is one of {\tt halph} (= \ha), {\tt bband} (= white-light) 
or {\tt caiik} (=\caii) and [yyyymmdd]\_[HHMMSS] is the date and time of 
image capture.

\subsubsection{Main archive data products}

The main archive contains all images of quality classes 1 and 2 in raw format
(only FITS header updated) and as quick look JPEG files in various types as 
described below. The cadence of the data is three images per minute 
for \ha\ and \caii\ observations and one image per minute for white-light. 
In case of flares this cadence is increased to 10 images per minute for 
chromospheric observations. On a day with good observing conditions and
a very active Sun up to 10\,000 images can be stored in the archive. 

\begin{itemize}
    \item[The archive consists of the following files in each daily directory]
    \item[]
    \item {\sf kanz\_[type]\_fi\_[yyyymmdd]\_[HHMMSS].fts.gz} 
      \begin{itemize}
        \item located in {\sf [yyyymmdd]/processed} 
        \item raw image files, only the FITS header is updated to its 
          full version (see Appendix Table \ref{tbl:FITS})
        \item example: \href{http://cesar.kso.ac.at/phokada/2017/20170906/processed/kanz_bband_fi_20170906_085054.fts.gz}{kanz\_bband\_fi\_20170906\_085054.fts.gz} 
        white-light image of 6~September~2017
      \end{itemize}
    \item {\sf kanz\_[type]\_fi\_[yyyymmdd]\_[HHMMSS].jpg} 
      \begin{itemize}
        \item located in {\sf [yyyymmdd]/JPEG} 
        \item greyscale quick look jpg files with annotation and 
           logos and unsharp mask applied for better contrast.
        \item example: \href{http://cesar.kso.ac.at/phokada/2017/20170906/JPEG/kanz_bband_fi_20170906_085054.jpg}{kanz\_bband\_fi\_20170906\_085054.jpg}
        white-light image of 6~September~2017
      \end{itemize}
    \item {\sf dc[yyyymmdd]\_[HHMMSS].fts.gz} 
      \begin{itemize}
        \item located in {\sf [yyyymmdd]/dark} 
        \item dark current images, normally two per day, at the beginning 
        and at the end of an observation day. The dark current is taken,
        when objective lens is covered with its protection cap.
        \item example: \href{http://cesar.kso.ac.at/phokada/2018/20180127/dark/dc20180127_073644.FTS.gz}{dc20180127\_073644.FTS.gz}
        for white-light camera of 27~January~2018
      \end{itemize}
    \item {\sf [yyyymmdd]\_[HHMMSS].jpg} 
      \begin{itemize}
        \item located in {\sf [yyyymmdd]/esa} 
        \item coloured JPEG files with logos, annotation and heliographic 
          grid and unsharp masking for contrast enhancement (not existing for \caii).
        \item example: \href{http://cesar.kso.ac.at/halpha3a/2021/20210422/esa/20210422_060537_fc.jpg}{20210422\_060537.jpg}
        \ha\ image of 22~April~2021
      \end{itemize}
    \item {\sf [yyyymmdd]\_[HHMMSS]\_fc.jpg} 
      \begin{itemize}
        \item located in {\sf [yyyymmdd]/esa}
        \item coloured jpg files with logos, annotation and heliographic grid and 
           corrected global anisotropies (not existing for \caii).
        \item example:\href{http://cesar.kso.ac.at/halpha3a/2021/20210422/esa/20210422_060537_fc.jpg}{20210422\_060537\_fc.jpg}
        \ha\ image of 22~April~2021
      \end{itemize}
    \item {\sf [yyyymmdd].avi} 
      \begin{itemize}
        \item located in {\sf [yyyymmdd]/}
        \item a daily overview movie generated from greyscale JPEG files.
        \item example: \href{http://cesar.kso.ac.at/caiia/2020/20201107/20201107.avi}{20201107.avi}
        \caii\ movie for 7~November~2020
      \end{itemize}
    \item {\sf [yyyymmdd]\_[HHMM]\_[IdId].html}
      \begin{itemize}
        \item located in {\sf [yyyymmdd]/movie}
        \item javascript movies of automatically detected flares. These movies
         are only produced for \ha\ flares.
        \item example: \href{http://cesar.kso.ac.at/halpha3a/2017/20170905/movie/20170905_0745_0003.html}{20170905\_0745\_0003.html}
        movie of importance class 1B from 5~September~2017
      \end{itemize}
    \item {\sf [yyyymmdd]\_movie.html} 
      \begin{itemize}
        \item located in {\sf [yyyymmdd]/}
        \item a daily javascript movie of coloured JPEG files. This movie 
        is only produced for \ha.
        \item example: \href{http://cesar.kso.ac.at/halpha3a/2014/20140127/20140127_movie.html}{20140127\_movie.html}
        \ha\ movie for 27~January~2014
      \end{itemize}
\end{itemize}
{\sf [IdId]} denotes a 4-digit internal number for identifying 
the flare region in the automatic flare detection system number. The 
colours in the coloured images are red for \ha\ and grey for 
white-light. The sizes of the jpg images are reduced to 1024$\times$1024 
pixels.

\begin{figure} 
 \centerline{\fbox{\includegraphics[width=1.0\textwidth,clip=]{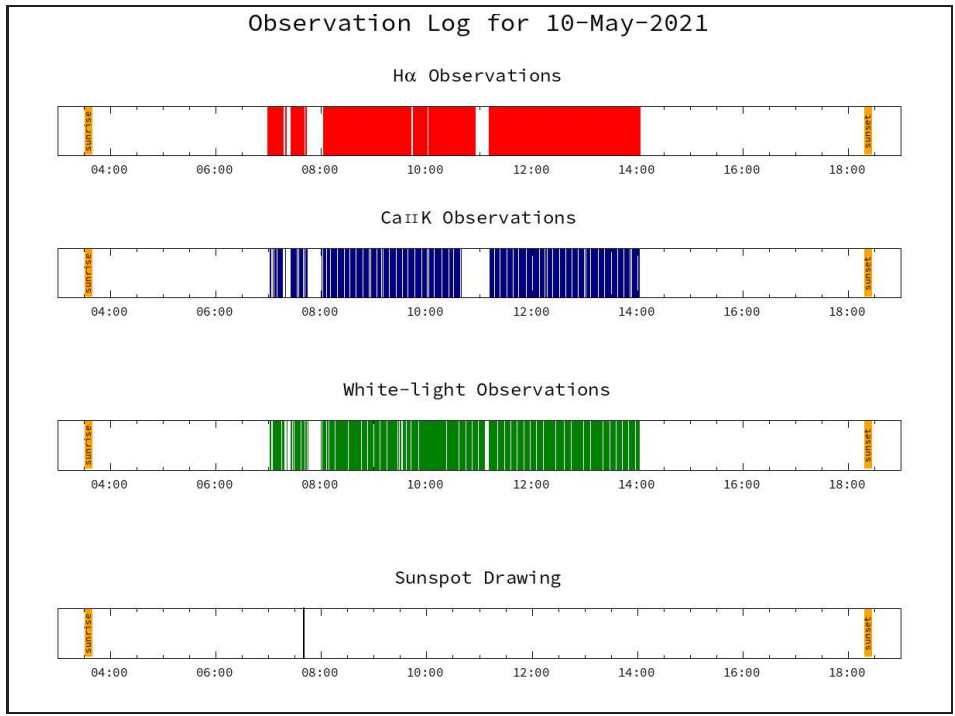}}}
 \caption{Graphical observation log for 10~May~2021. For each observed image
 of quality class 1 or 2 a line is drawn. Clouds at around 8:00 UT and 11:00 UT
 interrupted the observations.}
 \label{fig:obslog}
\end{figure}

\begin{figure} 
 \centerline{\includegraphics[width=1.0\textwidth,clip=]{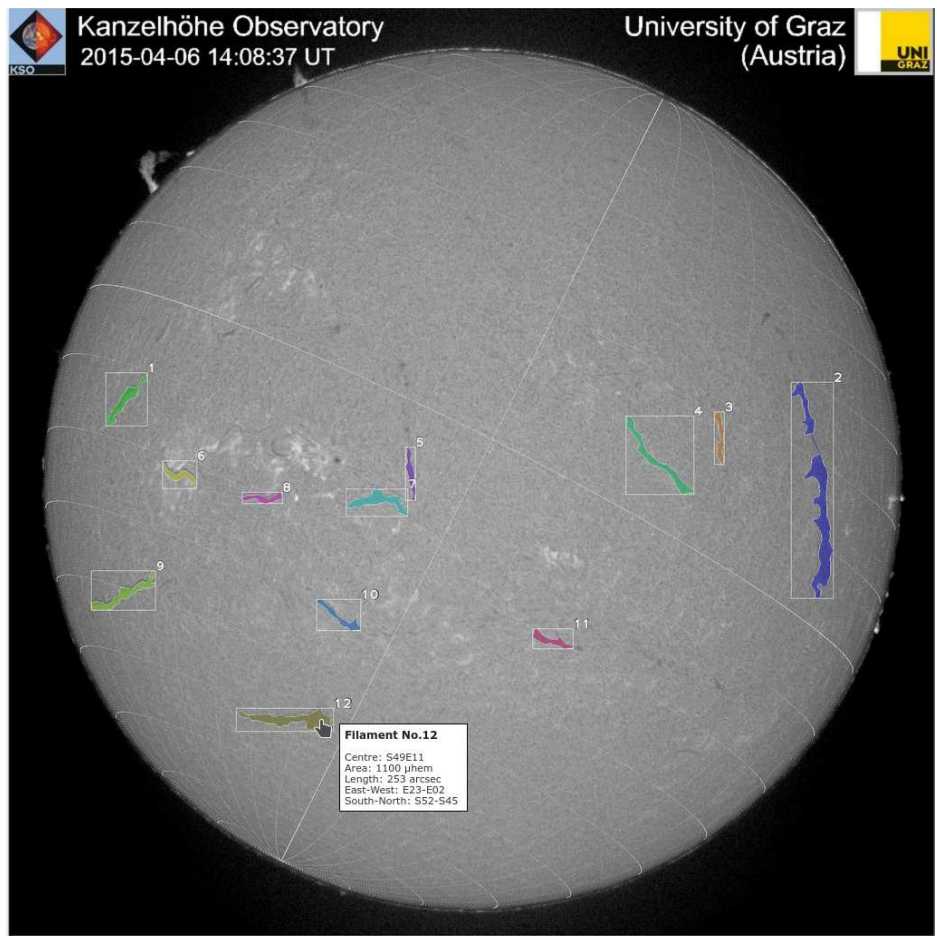}}
 \caption{\ha\ sensitive map of 6~April~2015: an \ha\ image is overlaid 
 with automatically detected filaments. When the mouse pointer is moved 
 over a filament additional information, like position and size, are shown.}
 \label{fig:fil}
\end{figure}

\subsubsection{Derived higher level data products}

\begin{figure} 
 \centerline{\includegraphics[width=1.0\textwidth,clip=]{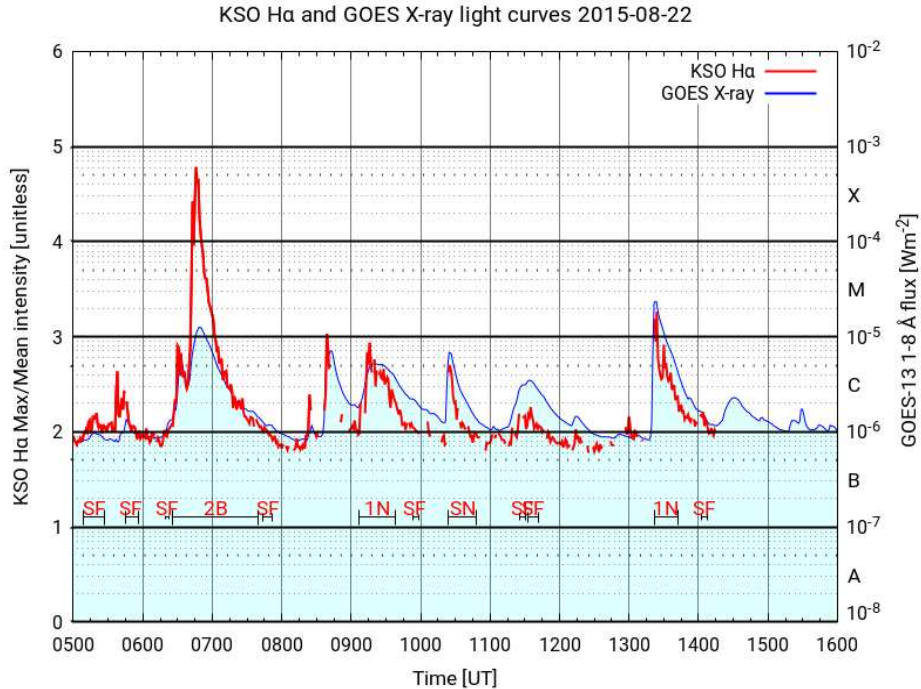}}
 \caption{The mean intensity of the four brightest pixels in each \ha\ 
 image normalized by the mean value of the full solar disc are plotted 
 in red along with the GOES 1-8 \AA\ X-ray flux in blue.  During flares 
 both curves should show a 
 local peak. The results of the automatic real-time flare detection 
 (flare start to end; \ha\ flare importance class) are plotted in the bottom.}
 \label{fig:lightc}
\end{figure}

The following higher level data products that are derived from the above data 
are  also available:
\begin{itemize}
\item Daily {\bf graphical observation logs} are created after each observing day
   to present a quick overview with time marks for each archived image
   for all cameras (see Figure \ref{fig:obslog}). 
   These are created after each observing day and made available via 
   the KSO archive webpage.
\item {\bf Sensitive filament maps} are generated from the filament data
   produced by the automatic flare and filament  detection system \citep{Poetzi2015}. 
   These filament data are combined to hourly sensitive maps (Figure \ref{fig:fil}),
   where \ha\ images are overlaid with detected filaments. When the 
   mouse pointer is moved over a filament, position, size and area 
   information is shown. Via the ESA SSA space weather portal
   also data from the database can be obtained in a JSON\footnote{JSON = 
   JavaScript Object Notation, \url{https://www.json.org/json-en.html}} format. 
\item {\bf Flare data and movies} which are also produced by the automatic
   flare detection system are available within the ESA SSA Space weather portal
   and the KSO archive webpage. Each detected \ha\ flare is characterized by 
   start time, peak time, end time and flare importance class. Each flare 
   is linked to its corresponding interactive javascript movie (described in the
   main archive products above).
\item In the {\bf \ha\ light curves} a plot of the brightest pixel in each 
   \ha\ image is overlaid by GOES full-Sun SXR flux profiles. The GOES data is 
   obtained from NOAA/SWPC (National Oceanic and Atmospheric 
   Administration/Space Weather Prediction Center), using the long 
   wavelength band (1.0 -- 8.0 \AA). Additionally automatically detected 
   flares together with their \ha\ importance class are annotated 
   in the plot (see Figure \ref{fig:lightc}). 
\end{itemize}

\subsubsection{Other data products and former data products}

The links to the different data products can be found in Tab. \ref{tbl:data}.

\begin{itemize}
\item {\bf Prominence images:} Once a day, when there are no thin clouds, 
  a set of \ha\ images is taken with the exposure times up to 50\,ms. 
  As a result the solar disc is completely saturated, but prominences 
  above the limb become better visible. Due to telescope limitations, prominences 
  can only be viewed up to 1.13 $R_{\odot}$. This data are used as 
  complementary data to fill gaps of the Lomnick\'y \v{S}t\'it prominence 
  catalogue \citep{Rybak2011}.
\item The daily {\bf sunspot drawings} are scanned immediately and available 
  on the KSO archive webpage together with the derived {\bf sunspot 
  relative numbers}. This is the longest data set of KSO reaching back to 1944.
\item {\bf \ha\ photographic film} data between 1973 and 2000 were scanned
  \citep{Poetzi2007} and are available as full FITS and JPEG archive.
\item {\bf \ha\ CCD} data observed with the former cameras (see Tab.\ref{tbl:patrol})
  with less resolution and less image depth are also available as synoptic
  and full data archive.
\item {\bf White-light photographic images}, of which normally three exposures on
  13$\times$18\,cm flat film were taken per day between 1989 and 2007, 
  have been scanned and processed \citep{Poetzi2010}. The archive consists 
  of the raw FITS files and JPEG images with overlaid heliographic grid.
\item {\bf NaD intensitygrams (5890 \AA), dopplergrams and magnetograms}
  of the Magneto Optical Filter (MOF) as described in \cite{Cacciani1999}
  \citep*[about the data reduction see][]{Moretti2000}.
  In the synoptic archive intensitygrams and magnetograms are available,
  the main archive consists of dopplergrams, intensitygrams and magnetograms.
  The data is only available from June~2000 to July~2002.
\end{itemize}

\begin{table}
 \caption{Summary of the links to KSO data products. (1 - web interface, 
  2 - to access the ESA-SSA data a free of charge registration is necessary.}
 \label{tbl:data}
 \begin{tabular}{ll} \hline
 data & link  \\ \hline
 \ha\ synoptic$^1$               &  \url{http://cesar.kso.ac.at/synoptic/ha_years.php} \\
 \ha\ 4 MPix \@ 12bit archive    &  \url{http://cesar.kso.ac.at/halpha3a} \\
 \caii\ synoptic$^1$             &  \url{http://cesar.kso.ac.at/synoptic/caii_years.php}  \\
 \caii\  archive                 &  \url{http://cesar.kso.ac.at/caiia}  \\
 white-light synoptic$^1$        &  \url{http://cesar.kso.ac.at/synoptic/con_years.php}  \\
 white-light archive             &  \url{http://cesar.kso.ac.at/phokada}  \\
 graphical observation logs$^1$  &  \url{http://cesar.kso.ac.at/kh_obslog/kh_obslog_query.php}  \\
 prominence images$^1$           &  \url{http://cesar.kso.ac.at/synoptic/kor_years.php}  \\
 sunspot drawings$^1$            &  \url{http://cesar.kso.ac.at/synoptic/draw_years.php} \\
 sunspot numbers$^1$             &  \url{http://cesar.kso.ac.at/synoptic/num_years.php} \\
 \ha\ scanned film archive       &  \url{http://cesar.kso.ac.at/hafilma}  \\
 \ha\ 1 MPix \@ 8bit archive     &  \url{http://cesar.kso.ac.at/halpha1a} \\
 \ha\ 1 MPix \@ 10bit archive    &  \url{http://cesar.kso.ac.at/halpha2a} \\
 MOF archive + synoptic          &  \url{http://cesar.kso.ac.at/mof} \\ 
 filament data$^1$               &  \url{http://cesar.kso.ac.at/sn_iv/filaments.php} \\
 \ha\ light curves ESA-SSA       &  \url{http://cesar.kso.ac.at/flares/intensity} \\ \\
 filament data ESA-SSA$^2$       &  \url{https://swe.ssa.esa.int/web/guest/kso-S107f-federated} \\
 flare data  ESA-SSA$^2$         &  \url{https://swe.ssa.esa.int/web/guest/kso-S107c-federated} \\
 \ha\ light curves ESA-SSA$^2$   &  \url{https://swe.ssa.esa.int/web/guest/kso-S107g-federated} \\
 \hline
 \end{tabular}
\end{table}

\subsection{KSO Database}\label{subDB}

The database is located on Kanzelh\"ohe Electronic Archive System 
\citep[KEAS,][]{otruba2007} and holds all compulsory
information about the images. The database is updated with every image 
transferred to the archive. {\it E.g.}, the following data is stored for each image:
\begin{itemize}
  \item the camera brand and CCD size
  \item the filter brand, wavelength and FWHM
  \item date and time of image acquisition
  \item exposure time in milliseconds 
  \item image quality (1-3)
  \item centre coordinates $x$ and $y$ of the solar disc in pixel
  \item radius of the solar disc in pixel 
  \item image size in $x$ and $y$ direction 
  \item filesize in bytes of the FITS file 
  \item filename of FITS file including path as accessible via  http
  \item available JPEG file types
\end{itemize}

\begin{figure} 
 \centerline{\fbox{\includegraphics[width=1.0\textwidth,clip=]{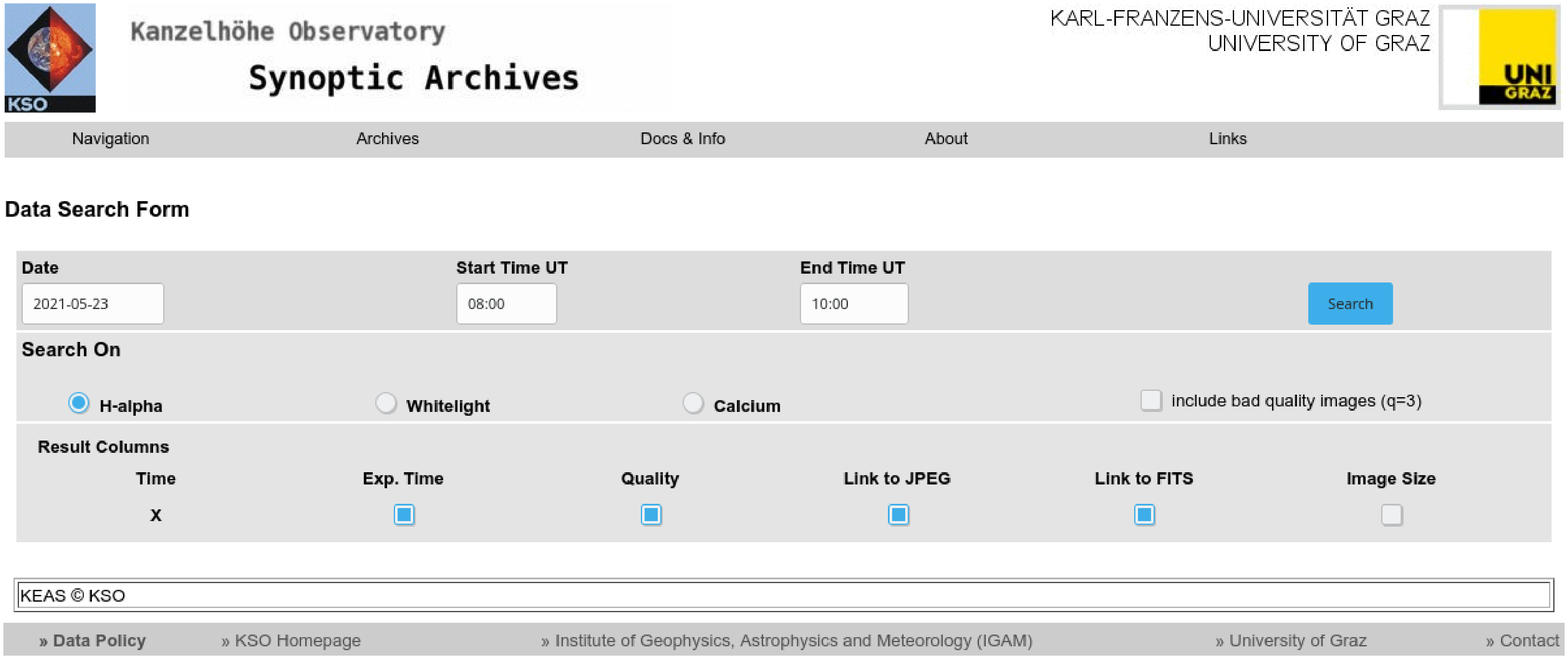}}}
 \centerline{\fbox{\includegraphics[width=1.0\textwidth,clip=]{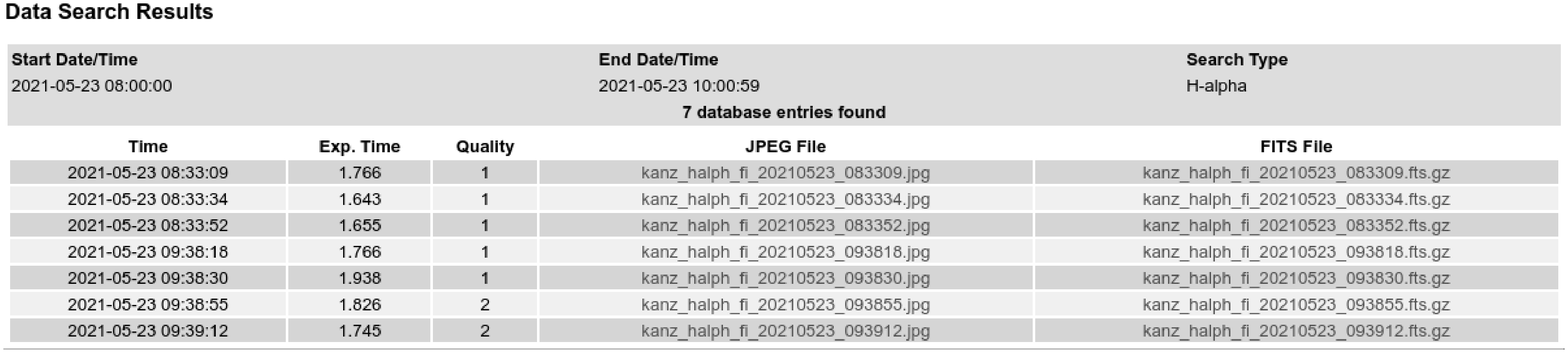}}}
 \caption{Example for the database query interface and result: Above the 
 database query for \ha\ images between 08:00 UT and 10:00 UT for 23~May~2021. 
 Below the result of this query, the file entries are links that can be viewed
 directly.}
 \label{fig:database}
\end{figure}

The database has nearly 5 million entries of solar images. All of them 
are accessible through the archive. Most entries are \ha\ images (about 3.5 million). 
For each entry there exists a FITS file and at least one JPEG file, 
in case of good quality additionally a coloured JPEG file and a coloured high contrast
JPEG file are available.
The image database can be accessed via the archive page directly
\url{http://cesar.kso.ac.at/catalogue/search_obj.php} (left side ``Database Search''). 
Figure \ref{fig:database} shows the query page and the result of this query beneath.

For scripts and automatized searches following APIs are available:\\
\url{http://cesar.kso.ac.at/database/get_halpha.php?from=datetime&to=datetime}\\
\url{http://cesar.kso.ac.at/database/get_wl.php?from=datetime&to=datetime}\\
\url{http://cesar.kso.ac.at/database/get_caii.php?from=datetime&to=datetime}\\
The parameters are the following:
\begin{itemize}
  \item {\sf from:} start date and time in the format {\sf yyyy-mm-ddTHH:MM:SS}.
  \item {\sf to:} end date and time in the format {\sf yyyy-mm-ddTHH:MM:SS}.
  \item {\sf ftype:} (optional) can be one of {\sf all, jpeg, fits}.
  \item {\sf color:} (optional) set to 1 if coloured normal contrast and
  high contrast (global anisotropies corrected) images are wanted.
  \item {\sf html:} (optional) set to 1 if html output is requested, the default
  output is a list with comma separated entries (csv).
  \item {\sf qual:} (optional) select the highest quality level to select,
  default is 3 {\it i.e.} 1,2 and 3 are selected.
  \item {\sf filesize:} (optional) set to 1 if the filesize in bytes should be 
  returned, this can be of interest for estimating the download time.
\end{itemize}
{\it E.g.}, to get all \ha\ image entries between 08:00:00 UT and 10:00:00 UT for
23~May~2021 including file sizes and coloured JPEGs the API looks like this:\\
\burl{http://cesar.kso.ac.at/database/get_halpha.php?from=2021-05-23T08:00:00&to=2021-05-23T10:00:00&html=1}
This gives in principle the same output as shown in the result in Figure \ref{fig:database}.
These APIs are also used, {\it e.g.}, to populate the KSO data of the Virtual
Solar Observatory (VSO, \cite{Davey2014}), where all entries of our database 
are replicated back to 1973.

\section{Conclusion}

This paper gives an overview of the data processing pipeline, the data
archive and the data products of the KSO solar observations. 
The Kanzelh\"ohe data pipeline could act as an example of how to process 
data fast and effectively. The time from data acquisition to data provision
is outstanding low, it can really be called ``near real-time'' data.
This fast processing is due to the three separated tasks -- acquisition, 
processing and provision --  running on independent servers.
The processing part, {\it e.g.}, is even implemented on two workstations
to reduce the workload, but can in case of a hardware failure run also on
one workstation alone. The data are never stored locally on camera 
controllers, all data beginning from data acquisition to processed data 
and temporary data are on network attached storages (NAS) that are 
accessible from all workstations. 
This separation of tasks on different workstations has the advantage that all 
steps can run in parallel, each step is just waiting for data produced by
the previous step.
The second important point is an almost immediate and easy access to the
observed data. For this purpose data in lower cadence (one image per minute)
is transferred immediately after processing to the archive and made 
accessible via the database. At the end of an observing day the archive
and the database are filled up with the rest of the data. The data can then
be retrieved via web interfaces or ftp access. Database queries can be
done directly over the archive web page or automatized via scripts using 
the described APIs.

Both, the fast data processing and the fast data availability, can only 
exist due to the automatized data processing pipeline. The basis for
this pipeline is the quality determination of the incoming data, that 
decides whether an image is dismissed or further processed. The provided 
data products serve as direct input to automated solar feature detection 
algorithms and allow to study the Sun in high-resolution and high-cadence.

\appendix

\section{Camera specifications}\label{camera}

The cameras are equipped with a ON Semiconductor KAI-4021 mono\-chrome CCD sensor.
The number of active pixels is 2048$\times$2048, each with a size of $7.4\,\mu\rm{m} 
\times 7.4\,\mu\rm{m}$, resulting in a total sensor size of $15.1\,\rm{mm} \times 15.1\,\rm{mm}$.
The peak quantum efficiency is 55\% at 475\,nm, {\it i.e.} the quantum efficiency for 
\caii\ (393\,nm) is 45\%, for white-light (545\,nm) $\approx$50\%, and for 
\ha\ (656\,nm) 30\%. The maximum data rate is 40 MHz, which limits the frame
rate to about 9 images per second in 12bit mode. This frame rate is reduced to 7 frames
per second in our software when frame selection is active. The sensor is a interline progressive scan
CCD, which reads out all pixels simultaneously. The number of dead pixels is
zero for the \ha\ camera, one for the white-light camera and three for the \caii\
camera. These dead pixels are replaced by the average intensity of the two neighbouring
horizontal pixels. Detailed information about the image sensor can be found at 
\url{https://www.onsemi.com/pdf/datasheet/kai-4021-d.pdf}.

\section{Quality parameters}\label{qualparam}

The parameters for determination of image quality of \\
\noindent \ha\ filtergrams:
\begin{itemize}
  \item[] high exposure times ($exp \ge18$\,ms): quality class 3
  \item[] rms of solar radius determination $>5$ pixel: quality class 3
  \item[] global quadrant intensity ratio $>1.08$: quality class 3
  \item[] $exp$ $>5$\,ms and rms of solar radius $>$ 3: quality class 3 
  \item[] image sharpness $<50$: quality class 2
  \item[] image sharpness $<25$: quality class 3
  \item[] mean intensity in AOI $\le 400$ or $\ge 1200$ counts: quality class 3 
\end{itemize}

\noindent  \caii\ filtergrams:
\begin{itemize}
  \item[] high exposure times ($exp \ge25$\,ms): quality class 3
  \item[] rms of solar radius determination $>4$ pixel: quality class 3
  \item[] global quadrant intensity ratio $>1.10$: quality class 3
  \item[] $exp$ $>15$\,ms and rms of solar radius $>$ 3: quality class 3 
  \item[] image sharpness $<22.5$: quality class 2
  \item[] image sharpness $<14$: quality class 3
  \item[] mean intensity in AOI $\le 500$ or $\ge 1200$ counts: quality class 3 
\end{itemize}

\noindent white-light images: 
\begin{itemize}
  \item[] high exposure times ($exp \ge12$\,ms): quality class 3
  \item[] rms of solar radius determination $>3$ pixel: quality class 3
  \item[] global quadrant intensity ratio $>1.10$: quality class 3
  \item[] $exp$ $>8$\,ms and rms of solar radius $>$ 2: quality class 3 
  \item[] image sharpness $<8$: quality class 2
  \item[] image sharpness $<5$: quality class 3
  \item[] mean intensity in AOI $\le 500$ or $\ge 2500$ counts: quality class 3 
\end{itemize}

The contrast in white-light images seems to be more sensitive to clouds, therefore the
limit for the exposure time setting is here very restrictive. The radius detection
is most accurate for white-light as the photosphere shows a very sharp intensity 
drop at the limb. This explains the higher tolerance for the rms of the 
radius determination in \ha\ as compared to white-light. The granulation in 
white-light images disappears immediately when the image is smoothed. Therefore
the limit for the image sharpness is set very low, {\it i.e.}, this parameter is not very
suitable for this image type.

\section{FITS header}\label{FITS}

\begin{table}
 \tiny
 \caption{The FITS keywords for an \ha\ image, \caii\ or white-light
 have the same keywords. Here an example for white-light from 13~January~2021
 at 08:00:39 UT.}
 \label{tbl:FITS}
 \begin{tabular}{rlll}     
 \hline
 0& SIMPLE &= T &/ file does conform to FITS standard \\
 1& BITPIX &= 16 &/ number of bits per data pixel \\
 2& NAXIS &= 2 &/ number of data axes \\
 3& NAXIS1 &= 2048 &\\
 4& NAXIS2 &= 2048 & \\
 5& EXTEND &= 'F ' &/ no extensions \\
 6& FILENAME &\multicolumn{2}{l}{= 'kanz\_halph\_fi\_20210113\_080039.fts.gz'} \\
 7& DATE &= '2021-01-13T08:00:42' &/ file creation date (YYYY-MM-DDThh:mm:ss UT) \\
 8& DATE-OBS &= '2021-01-13T08:00:39' &/ Date of observation \\
 9& DATE-BEG &= '2021-01-13T08:00:39' &/ Date of observation \\
 10& TIMESYS &= 'UTC ' &  \\
 11& OBSVTRY &\multicolumn{2}{l}{= 'Kanzelhoehe Observatory'}\\
 12& TELESCOP &= 'KHPI ' &  \\
 13& INSTRUME &= 'HA2 ' &  \\
 14& DETECTOR &= 'TM4200-6' &/ Camera Type \\
 15& OBJECT &= 'Full Sun' & \\
 16& FILTER &= 'Zeiss Lyot Halpha'  &\\
 17& WAVELNTH &= 6562.8 &/ [ANG], FWHM=0.7 [ANG] \\
 18& WAVEMIN &= 6562.45 & \\
 19& WAVEMAX &= 6563.15 &  \\
 20& EXP\_TIME &= 2.058 &/ Exposure Time [ms] \\
 21& XPOSURE &= 0.002058 &/ [s] \\
 22& BSCALE &= 1 &/ default scaling factor \\
 23& BZERO &= 32768 &/ offset data range to that of unsigned short \\
 24& BUNIT &= 'CCD COUNTS'  &\\
 25& DATAMIN &= 0 & \\
 26& DATAMEAN &= 539  &\\
 27& DATAMAX &= 657  &\\
 28& CTYPE1 &= 'SOLAR\_X '  &\\
 29& CTYPE2 &= 'SOLAR\_Y '  &\\
 30& CUNIT1 &= 'arcsec '  &\\
 31& CUNIT2 &= 'arcsec '  &\\
 32& CRPIX1 &= 1024.5 &/ [pix] \\
 33& CRPIX2 &= 1024.5 &/ [pix] \\
 34& CDELT1 &= 1.026318 &/ [arcsec/pix] \\
 35& CDELT2 &= 1.026318 &/ [arcsec/pix] \\
 36& CRVAL1 &= 1.838296 & \\
 37& CRVAL2 &= 1.425863 & \\
 38& ANGLE &= -4.725313 &/ [deg] \\
 39& CROTA1 &= 4.725313 &/ [deg] \\
 40& CENTER\_X &= 1022.709 &/ [pix] \\
 41& CENTER\_Y &= 1023.111 &/ [pix] \\
 42& SOLAR\_R &= 955.2872 &/ [pix] \\
 43& RSUN\_REF &= 6.9938E+08 &/ [m] \\
 44& RSUN\_ARC &= 980.4283 &/ [arcsec] \\
 45& SOLAR\_P0 &= -3.970842 &/ [deg] \\
 46& SOLAR\_B0 &= -4.385699 &/ [deg] \\
 47& CAR\_ROT &= 2239 & \\
 48& QUALITY &= 1 &/ image quality [1-3] \\
 49& OBS\_TYPE &= 'HALPH ' & \\
 50& OBS\_PROG &= 'HALPHA PATROL' &\\
 51& TYPE-DP &= 'ARCHIVE ' &/ Data Processing Type \\
 52& EXP\_MODE &= 0 &/ Exp. Mode (0=auto,1=dbl,2=fix,3=both) \\
 53& PRE\_INT &= 600 &/ Preselected PixInt in AOI \\
 54& A\_O\_INT &= '800,900,800,1100' &/ Rect. for PixInt [X0,Y0,X1,Y1] \\
 55& ORIGIN &\multicolumn{2}{l}{= 'KANZELHOEHE OBSERVATORY, A-9521 TREFFEN, AUSTRIA'} \\
 56& \multicolumn{3}{l}{COMMENT Orientation: N up, W right, first pix is left bottom} \\
 57& \multicolumn{3}{l}{HISTORY No intensity processing applied} \\
 58& \multicolumn{3}{l}{END} \\
 \hline
 \end{tabular}
 \end{table}

The FITS header keywords (see Tab. \ref{tbl:FITS}) used at KSO are mostly 
in accordance with the Metadata Definition for Solar Orbiter Science Data 
published by ESA \citep{DeGroof2019}. Most of the keywords are clear, but some
need an additional explanation and sevral keywords have the same value
or meaning are included for compatibility reasons with older software.
\begin{itemize}
    \item [DATE-OBS = DATE-BEG] Date and time of image capture.
    \item [DATE] Date and time of FITS creation.
    \item [EXP\_TIME, XPOSURE] both are the exposure times but the first is
    in milliseconds and the latter in seconds.
    \item [CDELT1, CDELT2] the angular aperture of pixels in arcsec, this angle
    refers to the viewing angle and not to heliographic coordinates
    \item [CRVAL1, CRVAL2] the value of CRPIX1 and CRPIX2 in arcsec, it is 
    for the horizontal coordinate (CRPIX1-CENTER\_X)$\times$CRDELT1.
    \item [ANGLE, CROTA1] The inclination of solar north clockwise and
    counter clockwise, it is the sum of SOLAR\_P0 and the camera tilt angle.
    \item [RSUN\_REF] The reference radius of the Sun in meters based on
    the IAU \citep{Prvsa2016} value for the photosphere and for the chromosphere
    transit measurements in \ha\ from KSO were taken, resulting in a thickness 
    of 3\,600 km.
    \item [RSUN\_ARC] The actual diameter of the Sun in arcsec depending on the
    distance of the sun and RSUN\_REF.
    \item [PRE\_INT] The requested intensity in the AOI for the frame selection
    procedure which defines the exposure time.
    \item [A\_O\_INT] the coordinates of the AOI rectangle.
\end{itemize}

\begin{acks}
This research has received financial support from the European 
Union’s Horizon 2020 research and innovation program under grant 
agreement No. 824135 (SOLARNET).
\end{acks}

\paragraph*{Disclosure of Potential Conflicts of Interests} 
The authors declare that they have no conflicts of interest.

\bibliographystyle{spr-mp-sola}
\tracingmacros=2
\bibliography{literature}

\end{article} 

\end{document}